\documentclass[11pt,draftclsnofoot,onecolumn]{IEEEtran}

\usepackage{cite}
\usepackage{graphicx}
\usepackage{psfrag}
\usepackage{subfigure}
\usepackage{url}
\usepackage{stfloats}
\usepackage{amsmath}
\usepackage{array}
\usepackage{amsmath}
\usepackage{amsfonts}
\usepackage{amssymb}
\usepackage{epsfig}
\usepackage{algorithm}

\usepackage{mathrsfs}
\usepackage{algpseudocode}

\newtheorem{theorem}{{Theorem}}
\newtheorem{lemma}[theorem]{{Lemma}}

\newtheorem{definition}{{Definition}}
\newtheorem{remark}{{Remark}}

\begin{document}
\title{Dynamic Conjectures in Random Access Networks Using Bio-inspired Learning}
\author{Yi~Su and~Mihaela~van der Schaar \\
Department of Electrical Engineering, UCLA} \maketitle

\begin{abstract}
Inspired by the biological entities' ability to achieve reciprocity
in the course of evolution, this paper considers a conjecture-based
distributed learning approach that enables autonomous nodes to
independently optimize their transmission probabilities in random
access networks. We model the interaction among multiple
self-interested nodes as a game. It is well-known that the Nash
equilibria in this game result in zero throughput for all the nodes
if they take myopic best-response, thereby leading to a network
collapse. This paper enables nodes to behave as intelligent entities
which can proactively gather information, form internal conjectures
on how their competitors would react to their actions, and update
their beliefs according to their local observations. In this way,
nodes are capable to autonomously ``learn" the behavior of their
competitors, optimize their own actions, and eventually cultivate
reciprocity in the random access network. To characterize the
steady-state outcome of this ``evolution", the conjectural
equilibrium is introduced. Inspired by the biological phenomena of
``derivative action" and ``gradient dynamics", two distributed
conjecture-based action update mechanisms are proposed to stabilize
the random access network. The sufficient conditions that guarantee
the proposed conjecture-based learning algorithms to converge are
derived. Moreover, it is analytically shown that all the achievable
operating points in the throughput region are essentially stable
conjectural equilibria corresponding to different conjectures. We
also investigate how the conjectural equilibrium can be selected in
heterogeneous networks and how the proposed methods can be extended
to ad-hoc networks. Numerical simulations verify that the system
performance significantly outperforms existing protocols, such as
IEEE 802.11 Distributed Coordination Function (DCF) protocol and
priority-based fair medium access control (P-MAC) protocol, in terms
of throughput, fairness, convergence, and stability.
\end{abstract}


\section{Introduction}
\IEEEPARstart Multi-user communication systems represent competitive
environments, where networked devices compete for the limited
available resources and wireless spectrum. Most of these devices are
autonomous, and must adapt to the surrounding environment in a
totally distributed and unsupervised manner. Recently, a number of
emerging approaches have been considered to better understand,
analyze, and characterize the dynamics of multi-user interactions
among communication devices using biologically-inspired methods
\cite{Haykin}-\cite{Editor_2}. The scientific rationale for this is
that, as communication networks expand in size, the network
``entities" grow in their diversity and ability to gather and
process information, and hence, networks will increasingly come to
resemble the models of interaction and self-organization of
biological systems.

It is well-known that many biological species exhibit various levels
of learning abilities, which enables them to survive and evolve in
the process of natural selection \cite{Gene}-\cite{learning_book}.
In particular, game theory has been used for a long time as a
descriptive tool for characterizing the interaction of biological
agents learning to improve their utility, e.g. the chance of
``survival" for the selfish genes or organisms
\cite{Gene}\cite{Smith}. Various learning models have been
developed, largely in response to observations by biologists about
animal and human behavior \cite{Gene}\cite{learning_book}. Several
particular dynamic adjustment processes have received specific
attention in the theory of learning and evolution. For example,
replicator dynamics models how the share of the population using a
certain survival strategy grows at a rate proportional to that
strategy's current payoff. In the partial best response dynamics, a
fixed portion of the population switches during each time period
from their current action to a best response to the aggregate
statistic of the population play in the previous period. As a
result, the cooperative or altruistic behavior may be favored and
reciprocity is therefore established in the course of evolution
\cite{Nowak}.


Several learning models have been applied to solve multi-user
interaction problems in both wireline and wireless network settings
\cite{Haykin}\cite{Friedman}-\cite{Fangwen}. For instance,
appropriate learning solutions are studied in distributed
environments consisting of agents with very limited information
about their opponents, such as the Internet \cite{Friedman}. A class
of no-regret learning algorithms is proposed in the stochastic game
framework to enable cognitive radio devices to learn from the
environment and efficiently utilize the spectrum resource
\cite{Haykin}. A reinforcement learning algorithm is proposed in the
repeated game setting to design power control in wireless ad-hoc
networks \cite{Long}, where it is shown that the learning dynamics
eventually converge to Nash equilibrium (NE) and achieve
satisfactory performance. A novel learning approach is proposed for
wireless users to dynamically and efficiently share spectrum
resources by considering the time-varying properties of their
traffic and channel conditions \cite{Fangwen}.

This paper is concerned with developing distributed learning
mechanisms in random access communication networks from not only the
biological, but also the game-theoretic perspective. It is
well-known that myopic selfish behavior is detrimental in random
access communication networks \cite{CSMACA_game}. To avoid a network
collapse and encourage cooperation, we adopt the conjecture-based
model introduced by Wellman and others \cite{Wellman}\cite{CE} and
enable the cognitive communication devices to build belief models
about how their competitors' reactions vary in response to their own
action changes. The belief functions of the wireless devices are
inspired by the evolutionary biological concept of reciprocity,
which refers to interaction mechanisms in which the emergence of
cooperative behavior is favored by the probability of future mutual
interactions \cite{Smith}\cite{Nowak}. Specifically, by deploying
such a behavior model, devices will no longer adopt myopic, selfish,
behaviors, but rather they will form beliefs about how their actions
will influence the responses of their competitors and, based on
these beliefs, they will try to maximize their own welfare. The
steady state of such a play among belief-forming devices can be
characterized as a conjectural equilibria (CE). At the equilibrium,
devices compensate for their lack of information by forming an
internal representation of the opponents' behavior and preferences,
and using these ``conjectured responses" in their personal
optimization program \cite{CE}. More importantly, we show that the
reciprocity among these self-interested devices can be sustained.



In particular, the main contributions of this paper are as follows.
First, to cultivate cooperation in random access networks, we enable
self-interested autonomous nodes to form independent linear beliefs
about how their rival actions vary as a function of their own
actions. Inspired by two biological phenomena, namely ``derivative
action" in biological motor control system
\cite{MIT_thesis}\cite{MIT_walking} and ``gradient dynamics" in
biological mutation \cite{gradient_1}-\cite{gradient_3}, we design
two simple distributed learning algorithms in which all the nodes'
beliefs and actions will be revised by observing the outcomes of
past mutual interaction over time. Both conjecture-based algorithms
require little information exchange among different nodes and the
internal computation for each node is very simple. For both
algorithms, we investigate the stability of different operating
points and derive sufficient conditions that guarantee their global
convergence, thereby establishing the connection between the dynamic
belief update procedures and the steady-state CE. We prove that all
the operating points in the throughput region are stable CE and
reciprocity can be eventually sustained via the proposed
bio-inspired evolution. We also provide an engineering
interpretation of the proposed bio-inspired design to clarify the
similarities and differences between the proposed algorithms and
existing protocols, e.g. the IEEE 802.11 DCF.

Second, we investigate the relationship between the parameter
initialization of beliefs and Pareto-efficiency of the achieved CE.
In the economic market context, it has been shown that adjustment
processes based on conjectures and individual optimization may
sometimes be driven to Pareto-optimality \cite{Learning_PB}. To the
best of our knowledge, this is the first attempt in investigating
the Pareto efficiency of the conjecture-based approach in
communication networks. Importantly, it is shown that, regardless of
the number of nodes, there always exist certain belief
configurations such that the proposed distributed bio-inspired
learning algorithms can operate arbitrarily close to the Pareto
boundary of the throughput region while approximately maintaining
the weighted fairness across the entire network. Our investigation
provides useful insights that help to define convergent dynamic
adaptation schemes that are apt to drive distributed random access
networks towards efficient, stable, and fair configurations.

The rest of this paper is organized as follows. Section II presents
the system model of random access networks, reviews the existing
game theoretic solutions, and introduces the concept of CE. Based on
the intuition gained from ``derivative action" and ``gradient
dynamics", Section III develops two simple distributed learning
algorithms in which nodes form dynamic conjectures and optimize
their actions based on their conjectures. The stability of different
CE and the condition of global convergence are established. This
section also shows that nodes' conjectures can be configured to
stably operate at any point that is arbitrarily close to the Pareto
frontier in throughput region. Section IV addresses the topics of
equilibrium selection in heterogeneous networks and presents
possible extension to ad-hoc networks. Numerical simulations are
provided in Section V to compare the proposed algorithms with the
IEEE 802.11 DCF protocol and P-MAC protocol. Conclusions are drawn
in Section VI.

\section{System Description and Conjectural Equilibrium}
In this section, we describe the system model of random access
networks and define the investigated random access game. We also
discuss the existing game-theoretic solutions and introduce the
concept of conjectural equilibrium.

\subsection{System Model of Random Access Networks}
Following \cite{Rvs_Num}\cite{Low_JSAC}, we model the interaction
among multiple autonomous wireless nodes in random access networks
as a random access game.

\begin{figure}
\centering
\includegraphics[width=0.5\textwidth]{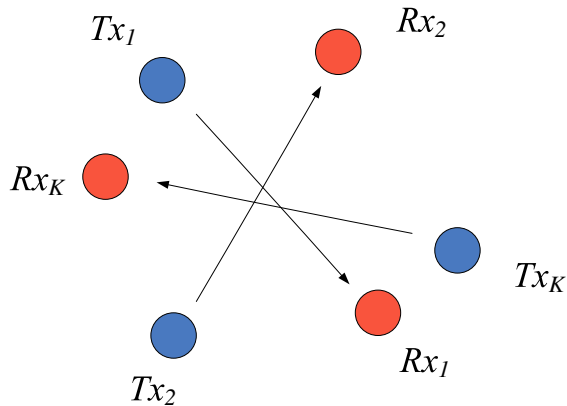}
\caption{System model of a single cell.}
\label{fg:onecell}
\end{figure}

As shown in Fig. \ref{fg:onecell}, consider a set
$\mathcal{K}=\{1,2,\ldots,K\}$ of wireless nodes and each node
represents a transmitter-receiver pair (link). We define $Tx_k$ as
the transmitter node of link $k$ and $Rx_k$ as the receiver node of
link $k$. We first assume a single-cell wireless network, where
every node can hear every other node in the network, and we will
address the ad-hoc network scenario in Section IV.B. The system
operates in discrete time with evenly spaced time slots
\cite{IT_throughput}\cite{Bell_lab}. We assume that all nodes always
have a data packet to transmit at each time slot (i.e. we
investigate the saturated traffic scenario\footnote{This paper
focuses on the saturated system because we are interested in
throughput maximization. The analysis can be extended to investigate
the non-saturated networks where the incoming packets of the
individual nodes' queues arrive at finite rates.}), and the network
is noise free and packet loss occurs only due to collision. The
action of a node in this game is to select its transmission
probability and a node $k$ will independently attempt transmission
of a packet with transmit probability $p_k$. The action set
available to node $k$ is $P_k=[0,1]$ for all
$k\in\mathcal{K}$\footnote{The action set can be alternatively
defined to be $P_k=[P_k^{\min},P_k^{\max}]$ and the analysis in this
paper still applies.}. Once the nodes decide their transmission
probabilities based on which they transmit their packets, an action
profile is determined. We denote the action profile in the random
access game as a vector $ \mathbf{p}=(p_1,\ldots,p_K)$ in
$P=P_1\times \cdots \times P_K$. Then the throughput of node $k$ is
given by\footnote{This throughput model assumes that time is slotted
and all packets are of equal length. We use this model for theoretic
analysis. The throughput of the scenarios in which packet lengths
are not equal, e.g. the IEEE 802.11 DCF, will be addressed in
Section V.}
\begin{equation}
\label{eq:eqn1} u_k(\mathbf{p})=p_k\prod_{i \in
\mathcal{K}\backslash\{k\}}(1-p_i).
\end{equation}
To capture the performance tradeoff in the network, the throughput
(payoff) region is defined as
$\mathscr{T}=\{(u_1(\textbf{p}),\ldots,u_K(\textbf{p}))| \ \exists \
\textbf{p}\in P\}$. The random access game can be formally defined
by the tuple $\Gamma=\langle\mathcal{K},(P_k),(u_k)\rangle$
\cite{Game_book}. Denote the transmission probability for all nodes
but $k$ by $ \mathbf{p}_{-k}=(p_1,\ldots,p_{k-1},p_{k+1},\\
\ldots,p_K)$. From (\ref{eq:eqn1}), we can see that node $k$'s
throughput depends not only on its own transmission probability
$p_k$, but also the other nodes' transmission probabilities
$\mathbf{p}_{-k}$.

\subsection{Existing Solutions}
The throughput tradeoff and stability of random access networks have
been extensively studied from the game theoretic perspective
\cite{Aloha_eq}-\cite{G_Aloha}. This subsection briefly reviews
these existing results and highlights the advantage and disadvantage
of different approaches.

In the random access game, one of the most investigated problems is
whether or not a Nash equilibrium exists. The definition of Nash
equilibrium is given as follows \cite{Game_book}.
\begin{definition}
A profile $ \mathbf{p}$ of actions constitutes a \emph{Nash
equilibrium} of $\Gamma$ if $u_k(p_k,\mathbf{p}_{-k})\geq
u_k(p'_k,\mathbf{p}_{-k})$ for all $p'_k\in P_k$ and
$k\in\mathcal{K}$.
\end{definition}

The NE of the investigated random access game has been addressed in
the similar context of CSMA/CA networks where selfish nodes
deliberately control their random deferment by altering their
contention windows \cite{CSMACA_game}. Specifically, the
transmission probability $p_k$ in our model can be related to the
contention window $CW_k$ in the CSMA/CA protocol, where
$p_k=\frac{2}{1+CW_k}$. It has been shown in \cite{CSMACA_game} that
at the NE, at least one selfish node will set $CW_k=1$ (i.e. always
transmit). If more than one selfish node sets its contention window
to 1, it will cause zero throughput for all the nodes in the system.
This kind of result is known as \emph{the tragedy of the commons}.
We can see that, myopic selfish behavior is detrimental in random
access scenarios and novel mechanisms are required to encourage
cooperative behavior among the self-interested devices. In addition,
the existence of and convergence to the NE in random access games
have been studied also in other scenarios, where individual nodes
have utility functions that are different from (\ref{eq:eqn1})
\cite{Aloha_eq}\cite{Rvs_Num}. For example, the nodes in
\cite{Aloha_eq} adjust their transmission probabilities in an
attempt to attain their desired throughputs. A local utility
function is found for exponential backoff-based MAC protocols, based
on which these protocols can be reverse-engineered in order to
stabilize the network \cite{Rvs_Num}. However, due to the inadequate
coordination or feedback mechanism in these protocols, Pareto
optimality of the throughput performance cannot be guaranteed.


Several recent works also investigate how to design new distributed
algorithms that provably converge to the Pareto boundary of the
network throughput region \cite{CSMACA_game}-\cite{Num_nomsg}. A
distributed protocol is proposed in \cite{CSMACA_game} to guide
multiple selfish nodes to a Pareto-optimal NE by including penalties
into their utility functions. However, the penalties must be
carefully chosen. In \cite{Mac_Num}, the utility maximization is
solved using the dual decomposition technique by enabling nodes to
cooperatively exchange coordination information among each other.
Furthermore, it is shown in \cite{Num_nomsg} that network utility
maximization in random access networks can be achieved without
real-time message passing among nodes. The key idea is to estimate
the other nodes' transmission probabilities from local observations,
which in fact increases the internal computational overhead of
individual nodes.

As discussed before, the goal of this paper is to design a simple
distributed random access algorithm that requires limited
information exchanges among nodes and also stabilizes the entire
network. More importantly, this algorithm should be capable of
achieving high efficiency and of differentiating among heterogeneous
nodes carrying various traffic classes with different quality of
service requirements. As we will show later, the game-theoretic
concept of conjectural equilibrium provides such an elegant
solution.

\subsection{Conjectural Equilibrium}
In game-theoretic analysis, conclusions about the reached equilibria
are based on assumptions about what knowledge the players possess.
For example, the standard NE strategy assumes that every player
believes that the other players' actions will not change at NE.
Therefore, it chooses to myopically maximize its immediate payoff
\cite{Game_book}. Therefore, the players operating at equilibrium
can be viewed as decision makers behaving optimally with respect to
their \emph{beliefs} about the strategies of other players.

To rigorously define CE, we need to include two new elements
$\mathcal{S}$ and $s$ and, based on this, reformulate the random
access game $\Gamma' =
\Big(\mathcal{K},(P_k),(u_k),(\mathcal{S}_k),(s_k)\Big)$
\cite{Wellman}. $\mathcal{S}=\times_{k\in \mathcal{K}}\mathcal{S}_k$
is the \emph{state space}, where $\mathcal{S}_k$ is the part of the
state relevant to the node $k$. Specifically, the state in the
random access game is defined as the contention probability that
nodes experience. The utility function $u_k$ is a map from the
nodes' state space to real numbers, $u_k:\mathcal{S}_k\times
P_k\rightarrow\mathcal{R}$. The \emph{state determination function}
$s=\times_{k\in \mathcal{K}}s_k$ maps joint action to state with
each component $s_k:P\rightarrow\mathcal{S}_k$. Each node cannot
directly observe the actions (transmission probabilities) chosen by
the others, and each node has some belief about the state that would
result from performing its available actions. The \emph{belief
function} $\tilde{s}_k$ is defined to be
$\tilde{s}_k:P_k\rightarrow\mathcal{S}_k$ such that
$\tilde{s}_k(p_k)$ represents the state that node $k$ believes it
would result in if it selects action $p_k$ . Notice that the beliefs
are not expressed in terms of other nodes' actions and preferences,
and the multi-user coupling in these beliefs is captured directly by
individual nodes forming conjectures of the effects of their own
actions. Moreover, each node chooses the action $p_k\in P_k$ if it
believes that this action will maximize its utility.

\begin{definition}
In the game $\Gamma'$ defined above, a configuration of belief
functions $(\tilde{s}_1^*,\ldots,\tilde{s}_K^*)$ and a joint action
$p^*=(p_1^*,\ldots,p_K^*)$ constitute a conjectural equilibrium, if
for each $k\in \mathcal{K}$,
\begin{displaymath} \tilde{s}_k^*(p_k^*)=s_k(p_1^*,\ldots,p_K^*) \ \textrm{and} \ p_k^*=\arg \max_{p_k\in P_k}u_k(\tilde{s}_k^*(p_k),p_k).
\end{displaymath}
\end{definition}

From the above definition, we can see that, at CE, all nodes'
expectations based on their beliefs are realized and each node
behaves optimally according to its expectation. In other words,
nodes' beliefs are consistent with the outcome of the play and they
behave optimally with respect to their beliefs. The key challenges
are how to configure the belief functions such that reciprocal
behavior is encouraged and how to design the evolution rules such
that the network can dynamically converge to a CE having
satisfactory performance. Section III provides bio-inspired
solutions for these problems in random access games.
\section{Distributed Bio-inspired Learning}
In this section, to promote reciprocity, we design a prescribed rule
for each node to configure its belief about its expected contention
of the wireless network as a linear function of its own transmission
probability. It is shown that all the achievable operating points in
the throughput region $\mathscr{T}$ are CE by deploying these belief
functions. Furthermore, inspired by the biological mechanisms
``derivative action" and ``gradient dynamics", we propose two
distributed learning algorithms for these nodes to dynamically
achieve the CE. We provide the sufficient conditions that guarantee
the stability and convergence of the CE. We also discuss the
similarities and differences between these bio-inspired algorithms
and the existing well-known protocols. Finally, it is proven that
any Pareto-inefficient operating point is a stable CE, i.e. we can
approach arbitrarily close to the Pareto frontier of the throughput
region $\mathscr{T}$.
\subsection{Individual Behavior}
As discussed before, both the state space and belief functions need
to be defined in order to investigate the existence of CE. In the
random access game, we define the state $s_k=\prod_{i \in
\mathcal{K}\backslash\{k\}}(1-p_i)$ to be the contention measure
signal representing the probability that all nodes except node $k$
do not transmit. This is because besides its own transmission
probability, its throughput only depends on the probability that the
remaining nodes do not transmit. We can see that state $s_k$
indicates the aggregate effects of the other nodes' joint actions on
node $k$'s payoff. In practice, it is hard for wireless nodes to
compute the exact transmission probabilities of their opponents
\cite{Num_nomsg}. Therefore, we assume that $s_k$ is the only
information that node $k$ has about the contention level of the
entire network, because it is a metric that node $k$ can easily
compute based on local observations. Specifically, from user $k$'s
viewpoint, the probabilities of experiencing an idle time slot is
$p_k^{idle}=(1-p_k)s_k$. Let $n_k^{idle}$ denote the number of time
slots between any two consecutive idle time slots. $n_k^{idle}$ has
an independent identically distributed geometric distribution with
probability $p_k^{idle}$. Therefore, we have
$p_k^{idle}=1/(1+\bar{n}_k^{idle})$, where $\bar{n}_k^{idle}$ is the
mean value of $n_k^{idle}$ and can be locally estimated by node $k$
through its observation of the channel contention history. Since
node $k$ knows its own transmission probability $p_k$, it can
estimate $s_k$ using $s_k=1/(1+\bar{n}_k^{idle})(1-p_k)$. Notice
that the action available to node $k$ is to choose the transmission
probability $p_k\in P_k$. By the definition of belief function, we
need to express the expected contention measure $\tilde{s}_k$ as a
function of its own transmission probability $p_k$. The simplest
approach is to deploy linear belief models, i.e. node $k$'s belief
function takes the form
\begin{equation}
\label{eq:eqn3} \tilde{s}_k(p_k)=\bar{s}_k-a_k(p_k-\bar{p}_k),
\end{equation}
for $k\in \mathcal{K}$. The values of $\bar{s}_k$ and $\bar{p}_k$
are specific states and actions, called \emph{reference points}
\cite{Learning_PB} and $a_k$ is a positive scalar. In other words,
node $k$ assumes that other nodes will observe its deviation from
its reference point $\bar{p}_k$ and the aggregate contention
probability deviates from the referent point $\bar{s}_k$ by a
quantity proportional to the deviation of $p_k-\bar{p}_k$. How to
configure $\bar{s}_k, \bar{p}_k$, and $a_k$ will be addressed in the
rest of this paper. The reasons why we focus on the linear beliefs
represented in (\ref{eq:eqn3}) are two-fold. First, the linear form
represents the simplest model based on which a user can model the
impact of its environment. As we will show later in Section III-E,
building and optimizing over such simple beliefs is sufficient for
the network to achieve almost any operating point in the throughput
region as a stable CE. Second, the conjecture functions deployed by
the wireless users are based on the concept of reciprocity
\cite{Smith}\cite{Nowak}, which was developed in evolutionary
biology, and refers to interaction mechanisms in which the evolution
of cooperative behavior is favored by the probability of future
mutual interactions. Similarly, in single-hop wireless networks, the
devices repeatedly interact when accessing the channel. If they
disregard the fact that they have a high probability to interact in
the future, they will act myopically, which will lead to a tragedy
of commons (the zero-payoff Nash equilibrium). However, if they
recognize that their probability of interacting in the future is
high, they will consider their impact on the network state, which is
captured in the belief function by the positive $a_k$.

The goal of node $k$ is to maximize its expected throughput
 $p_k\cdot\tilde{s}_k(p_k)$ taking into account the conjectures that it
has made about the other nodes. Therefore, the optimization a node
needs to solve becomes:
\begin{equation}
\label{eq:eqn4} \max_{p_k\in
P_k}p_k\Big[\bar{s}_k-a_k(p_k-\bar{p}_k)\Big],
\end{equation}
where the second term is the expected contention measure
$\tilde{s}_k(p_k)$ if node $k$ transmits with probability $p_k$. The
product of $p_k$ and $\tilde{s}_k(p_k)$ gives the expected
throughput for $p_k\in P_k$. For $a_k>0$, node $k$ believes that
increasing its transmission probability will increase its
experienced contention probability. The optimal solution of
(\ref{eq:eqn4}) is given by
\begin{equation}
\label{eq:eqn5} p_k^*= \min \Big\{ \frac
{\bar{s}_k}{2a_k}+\frac{\bar{p}_k}{2} ,1 \Big\}.
\end{equation}

In the following, we first show that forming simple linear beliefs
in (\ref{eq:eqn3}) can cause all the operating points in the
achievable throughput region to be CE.
\begin{theorem}
\label{th:th1} All the operating points in the throughput region
$\mathscr{T}$ are conjectural equilibria.
\end{theorem}

\emph{Proof}: For each operating point $(\tau_1,\ldots,\tau_K)$ in
the throughput region $\mathscr{T}$, there exists at least a joint
action profile $(p^*_1,\ldots,p^*_K)\in P$ such that
$\tau_k=u_k(\textbf{p}^*)$, $\forall k\in\mathcal{K}$. We consider
setting the parameters in the belief functions to be:
\begin{equation}
\label{eq:eqn8} a_k^*=\frac{\prod_{i \in
\mathcal{K}\backslash\{k\}}(1-p_i^*)}{p_k^*}.
\end{equation}
It is easy to check that, if the reference points are
$\bar{s}_k=\prod_{i
\in\mathcal{K}\backslash\{k\}}(1-p_i^*),\bar{p}_k=p_k^*$, we have
$\tilde{s}_k(p_k^*)=s_k(p_1^*,\ldots,p_K^*)$ and $p_k^*=\arg
\max_{p_k\in P_k}u_k(\tilde{s}_k(p_k),p_k)$. Therefore, this
configuration of the belief functions and the joint action
$\textbf{p}^*=(p_1^*,\ldots,p_K^*)$ constitute the CE that results
in the throughput $(\tau_1,\ldots,\tau_K)$.\footnote{By the
definition of CE, the configuration of the linear belief functions
is a key part of CE. Since this paper focuses on the linear belief
functions defined in (\ref{eq:eqn3}), we will simply state the joint
action $\textbf{p}^*$ is a CE hereafter for the ease of
presentation.} $\blacksquare$

Theorem \ref{th:th1} establishes the existence of CE, i.e. for a
particular $\textbf{p}^*\in P$, how to choose the parameters
$\{\bar{s}_k,\bar{p}_k,a_k\}_{k=1}^K$ such that $\textbf{p}^*$ is a
CE. However, it neither tells us how these CE can be achieved and
sustained in the dynamic setting nor clarifies how different belief
configurations can result in various CE.

In distributed learning scenarios, nodes learn when they modify
their conjectures based on their new observations. Specifically, we
first allow the nodes to revise their reference points based on
their past local observations. Let
$s_k^t,p_k^t,\tilde{s}_k^t,\bar{s}_k^t,\bar{p}_k^t$ be user $k$'s
state, transmission probability, belief function, and reference
points at stage $t$\footnote{This paper assumes the persistence
mechanism for contention resolution except in Section III.D. In the
persistence mechanism, each wireless node maintains a persistence
probability and accesses the channel with this probability
\cite{Mobicom_ref}. A stage contains multiple time slots. The nodes
estimate the contention level in the network and update their
persistence probabilities in the "stage-by-stage" manner. The
superscript $t$ in this paper represents the numbering of the stages
unless specified.}, in which $s_k^t=\prod_{i \in
\mathcal{K}\backslash\{k\}}(1-p_i^t)$. We propose a simple rule for
individual nodes to update their reference points. At stage $t$,
node $k$ set its $\bar{s}_k^t$ and $\bar{p}_k^t$ to be $s_k^{t-1}$
and $p_k^{t-1}$. In other words, node $k$'s conjectured utility
function at stage $t$ is
\begin{equation}
\label{eq:eqn6} u_k^t(\tilde{s}_k^t(p_k),p_k)=p_k\Bigl[\prod_{i \in
\mathcal{K}\backslash\{k\}}(1-p_i^{t-1})-a_k(p_k-p_k^{t-1})\Bigr].
\end{equation}
The remainder of this paper will investigate the dynamic properties
of the resulting operating points and the performance trade-off
among multiple competing nodes. In particular, for fixed
$\{a_k\}_{k=1}^K$, Sections III-B and C will embed the above
individual optimization scheme in two different distributed learning
processes in which all the nodes update their transmission
probabilities over time. Section III-E further allows individual
nodes adaptively update their parameters $\{a_k\}_{k=1}^K$ such that
desired efficiency can be attained. For given $\{a_k\}_{k=1}^K$,
Section IV-A will derive a quantitative description of the resulting
CE $\textbf{p}^*$.

\subsection{A Best Response Learning Algorithm}
Our first algorithm in establishing reciprocity through a evolution
process is inspired by the ``derivative action", which is a key
component of biological motor control system models, e.g. cerebellar
control over arm, hand, truncal, and leg movements
\cite{MIT_thesis}\cite{MIT_walking}. Specifically, during limb
movements, high frequency differential (velocity-like) signals after
filtering due to biological sensors are attributed to lateral
cerebellum as part of the input for cerebellar control. The
classical control interpretation of ``derivative action" is that the
first-order derivative term serves as a short term prediction of the
measured zero-order variable. For example, in a swing leg control,
the velocity-like signals enable a cerebrocerebellar channel to
better locate the ankle (or foot) position in front of the hip
position during the swing phase. The conjecture-based approach is
very similar in spirit with the aforementioned biological motor
control models and the standard proportional integral derivative
(PID) controllers in engineered systems \cite{PID}. In particular,
the derivative term is substituted by a node's internal belief of
how its own action will impact the other nodes' behavior. Our first
learning algorithm adopts the simplest update mechanism in which
each node adjusts its transmission probability using the best
response that maximizes its conjectured utility function
(\ref{eq:eqn6}). Therefore, at stage $t$, node $k$ chooses a
transmission probability
\begin{equation}
\label{eq:eqn7} p_k^t=\arg \max_{p_k\in P_k}
u_k^t(\tilde{s}_k^t(p_k),p_k)= \min \Big\{\frac{p_k^{t-1}}{2}+\frac
{\prod_{i \in
\mathcal{K}\backslash\{k\}}(1-p_i^{t-1})}{2a_k},1\Big\}.
\end{equation}
In this regard, the use of ``derivative action" by an agent can be
interpreted as using the best response to the forecasted effect of
all the opponents' strategies.

\begin{algorithm}
\caption{: A Distributed Best Response Learning Algorithm for Random
Access}
\begin{algorithmic}[1]
\State Initialize: $t=0$, the transmission probability
$p_k^0\in[0,1]$, and the parameter $a_k>0$ in node $k$'s belief
function, $\forall k\in\mathcal{K}$. \Procedure {}{} \State
\emph{Locally} at each node $k$, iterate through $t$: \State Set $t
\gets t+1.$ \ForAll {$k\in\mathcal{K}$} \State At stage $t$,
$p_k^t\gets \min \{p_k^{t-1}/2+\prod_{i \in
\mathcal{K}\backslash\{k\}}(1-p_i^{t-1})/(2a_k),1\}$. \EndFor \State
Node $k$ decides if it will transmit data with a
 probability $p_k^t$ (or equivalently, maintain a window size of $CW_k^t=2/p_k^t-1$) for all the time slots during stage $t$. \EndProcedure
\end{algorithmic}
\end{algorithm}

\begin{figure}
\centering
\includegraphics[width=0.8\textwidth]{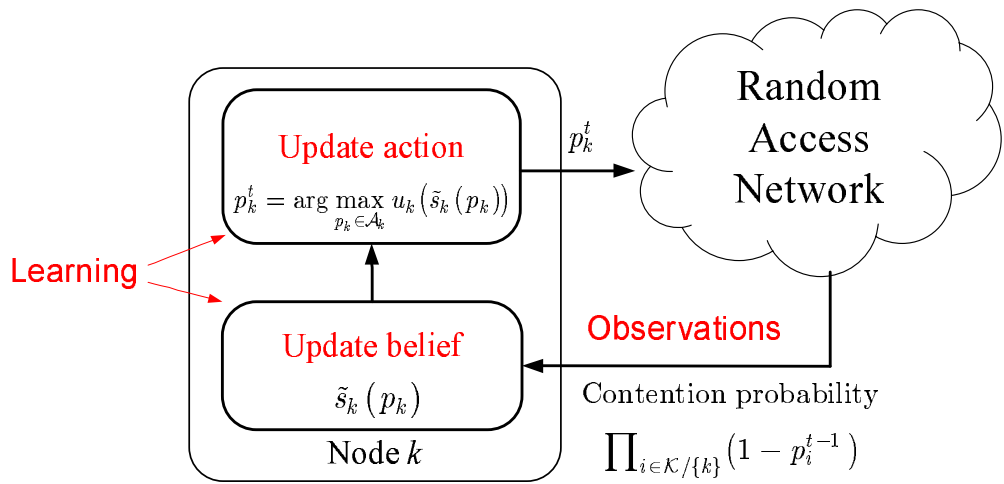}
\caption{An illustration of the distributed learning process.}
\label{fg:BR}
\end{figure}

The detailed description of the entire distributed best response
learning procedure is summarized in Algorithm 1 and it is also
pictorially illustrated in Fig. \ref{fg:BR}. Next, we are interested
in deriving the limiting behavior, e.g. stability and convergence,
 of this algorithm. For ease of illustration, the sufficient conditions for
 stability and convergence throughout this paper are expressed in terms of $\{p_k\}_{k=1}^K$
 and $\{a_k\}_{k=1}^K$, respectively. The mapping from
 $\{p_k\}_{k=1}^K$ to $\{a_k\}_{k=1}^K$ is given in (\ref{eq:eqn8}) and the mapping
 from $\{a_k\}_{k=1}^K$ to $\{p_k\}_{k=1}^K$ will be addressed in
 Section IV-A.

\subsubsection{Local Stability}

Although Theorem \ref{th:th1} indicates that all the points in
$\mathscr{T}$ are CE, they may not be necessarily stable. An
unstable equilibrium is not desirable, because any small
perturbation might cause the sequence of iterates to move away from
the initial equilibrium. The following theorem describes a subset in
$P$ in which all the points are stable CE.
\begin{theorem} \label{th:th2}
For any $\textbf{p}^*=(p_1^*,\ldots,p_K^*)\in P$, if
\begin{equation} \label{eq:eqn9}
\sum_{k=1}^{K}p_k^*<1, \quad \textrm{or} \quad \sum_{i \in
\mathcal{K}\backslash\{k\}}\frac{p_k^*}{1-p_i^*}<1, \ \forall k \in
\mathcal{K},
\end{equation} $\textbf{p}^*$ is a stable CE for Algorithm 1.
\end{theorem}

\emph{Proof}: To analyze the stability of different CE, we consider
the Jacobian matrix of the self-mapping function in (\ref{eq:eqn7}).
Let $J_{ik}$ denote the element at row $i$ and column $k$ of the
Jacobian matrix $\textbf{J}$. If $p_k^{t-1}/2+\prod_{i \in
\mathcal{K}\backslash\{k\}}(1-p_i^{t-1})/(2a_k)\leq1$, the Jacobian
matrix $\textbf{J}^{BR}$ of (\ref{eq:eqn7}) is defined as:
\begin{equation}
\label{eq:eqn10}
J^{BR}_{ik}=\frac{\partial p_i^t}{\partial p_k^{t-1}}=\left\{
\begin{array}{cl}
\frac{1}{2},& \text{if $i=k$},\\
-\frac{1}{2a_i}\prod_{j \in
\mathcal{K}\backslash\{i,k\}}(1-p_j^{t-1}),& \text{if $i\neq
k$}.\end{array} \right.
\end{equation}
As proven in Theorem \ref{th:th1}, for
$\textbf{p}^*=(p_1^*,\ldots,p_K^*)\in P $ to be a fixed point of the
self-mapping function in (\ref{eq:eqn7}), $a_k$ must be set to be
$a_k^*=\prod_{i \in \mathcal{K}\backslash\{k\}}(1-p_i^*)/p_k^*$. It
follows that
\begin{equation} \label{eq:eqn11}
J_{ik}^{BR}\rvert_{\textbf{p}=\textbf{p}^*,\
\textbf{\textit{a}}=\textbf{\textit{a}}^*}=\left\{
\begin{array}{cl}
\frac{1}{2},& \text{if $i=k$},\\
-\frac{p_i^*}{2(1-p_k^*)},& \text{if $i\neq k$}.
\end{array} \right.
\end{equation}
$\textbf{p}^*$ is stable if and only if the eigenvalues
$\{\lambda_k\}_{k=1}^{K}$ of matrix $\textbf{J}^{BR}$ in
(\ref{eq:eqn11}) are all inside the unit circle of the complex
plane, i.e. $|\lambda_k|<1, \forall k \in \mathcal{K}$.

From Gersgorin circle theorem \cite{Matrix_book}, all the
eigenvalues $\{\lambda_k\}_{k=1}^{K}$ of $\textbf{J}^{BR}$ are
located in the region
\begin{displaymath}
\bigcup_{k=1}^{K} \Big\{ |\lambda-J_{kk}^{BR}|\leqslant\sum_{i \in
\mathcal{K}\backslash\{k\}}|J_{ik}^{BR}| \Big\} \ \textrm{and} \
\bigcup_{k=1}^{K} \Big\{ |\lambda-J_{kk}^{BR}|\leqslant\sum_{i \in
\mathcal{K}\backslash\{k\}}|J_{ki}^{BR}| \Big\}.
\end{displaymath}
Note that $J_{kk}^{BR}=1/2$, these regions can be further simplified
as
\begin{displaymath} \bigcup_{k=1}^{K} \Big\{ |\lambda-\frac{1}{2}|\leqslant\sum_{i \in
\mathcal{K}\backslash\{k\}}\frac{p_i^*}{2(1-p_k^*)} \Big\}  \
\textrm{and} \ \bigcup_{k=1}^{K} \Big\{
|\lambda-\frac{1}{2}|\leqslant\sum_{i \in
\mathcal{K}\backslash\{k\}}\frac{p_k^*}{2(1-p_i^*)} \Big\}.
\end{displaymath}
If either condition in (\ref{eq:eqn9}) is satisfied, all the
eigenvalues of $\textbf{J}^{BR}$ must fall into the region
$|\lambda-\frac{1}{2}|<\frac{1}{2}$, which is located within the
unit circle $|\lambda|<1$. Therefore, $\textbf{p}^*$ is a stable CE.
\ $\blacksquare$



\begin{remark} \label{rm:rm2}
$p_k^*/(1-p_i^*)$ can be interpreted as the worst case probability
that node $k$ occupies the channel given that node $i$ does not
transmit. This metric reflects from node $k$'s perspective the
impact that node $i$'s evacuation has on the overall congestion of
the channel. Therefore, the sufficient conditions in (\ref{eq:eqn9})
means that if the system is not overcrowded from all the nodes'
perspectives, the corresponding CE is stable. We can see from
Theorem \ref{th:th2} that lowering the transmission probabilities
helps to stabilize the random access network. The system can
accommodate a certain degree of individual nodes' ``aggressiveness"
while maintaining the network stability. For example, if a node
sends its packets with a probability close to 1, as long as the
other nodes are conservative and they set their transmission
probability small enough, the entire network can still be
stabilized. However, if too many ``aggressive" nodes with large
transmission probabilities coexist, the system stability may
collapse, leading to a tragedy of commons.
\end{remark}

\subsubsection{Global Convergence}
Note that Theorem \ref{th:th2} only investigates the stability for
different fixed points, i.e. Algorithm 1 converges to these points
when initial values are close enough to them. In addition to local
stability, we are also interested in characterizing the global
convergence of Algorithm 1 when using various $a_k$ to initialize
the belief function $\tilde{s}_k$.
\begin{theorem}\label{th:th4}
Regardless of any initial value chosen for $\{p_k^0\}_{k=1}^{K}$, if
the parameters $\{a_k\}_{k=1}^{K}$ in the belief functions
$\{\tilde{s}_k\}_{k=1}^{K}$ satisfy
\begin{equation} \label{eq:eqn12}
\sum_{i \in \mathcal{K}\backslash\{k\}}\frac{1}{a_i}<1, \forall k
\in \mathcal{K},
\end{equation}
Algorithm 1 converges to a unique CE.
\end{theorem}

\emph{Proof}: For $a_k>1$, the self-mapping function in
(\ref{eq:eqn7}) can be rewritten as
\begin{equation} \label{eq:eqn13}
p_k^t=\frac{p_k^{t-1}}{2}+\frac {\prod_{i \in
\mathcal{K}\backslash\{k\}}(1-p_i^{t-1})}{2a_k}.\end{equation} We
can prove for Algorithm 1 the uniqueness of and the convergence to
CE by showing that function (\ref{eq:eqn13}) is a contraction map if
the condition in (\ref{eq:eqn12}) is satisfied.

Let $d(\cdot)$ be the induced distance function by certain vector
norm in the Euclidean space. Consider two sequences of the
transmission probability vectors $\{\textbf{p}^0,\ldots,
\textbf{p}^{t-1},\textbf{p}^t,\ldots\}$ and
$\{\hat{\textbf{p}}^0,\ldots,
\hat{\textbf{p}}^{t-1},\hat{\textbf{p}}^t,\ldots\}$. We have
\begin{equation} \label{eq:eqn14}
d(\textbf{p}^t,\hat{\textbf{p}}^t)=\lVert\textbf{p}^t-\hat{\textbf{p}}^t
\rVert\leq\lVert\textbf{J}^{BR}\rVert\cdot\lVert\textbf{p}^{t-1}-\hat{\textbf{p}}^{t-1}
\rVert=\lVert\textbf{J}^{BR}\rVert\cdot
d(\textbf{p}^{t-1},\hat{\textbf{p}}^{t-1}).
\end{equation}
The matrix norm used here is induced by the same vector norm. Using
$\lVert\cdot\rVert_1$ for the Jacobian matrix of (\ref{eq:eqn13}) as
given in (\ref{eq:eqn11}), we have
\begin{equation} \label{eq:eqn16}
\lVert\textbf{J}^{BR}\rVert_1=\max_{k \in
\mathcal{K}}\sum_{i=1}^K\lvert J_{ik}^{BR}\lvert \
\leq\frac{1}{2}+\frac{1}{2}\max_{k \in \mathcal{K}}\sum_{i \in
\mathcal{K}\backslash\{k\}}\frac{1}{a_i}.
\end{equation}
Therefore, if the condition in (\ref{eq:eqn12}) is satisfied, there
exist a constant $q\in[0,1)$ and a positive $\epsilon$, such that
$q=\lVert\textbf{J}^{BR}\rVert_1=1-\epsilon<1$ and
$\lVert\textbf{p}^t-\hat{\textbf{p}}^t \rVert_1\leq q
\lVert\textbf{p}^{t-1}-\hat{\textbf{p}}^{t-1} \rVert_1$. From the
contraction mapping theorem \cite{Contraction_mapping}, the
self-mapping function in (\ref{eq:eqn7}) has a unique fixed point
and the sequence $\{\textbf{p}^t\}_{t=0}^{+\infty}$ converges to the
unique fixed point. \ $\blacksquare$

\begin{remark} We can also alternatively derive a sufficient
condition using $\lVert\cdot\rVert_{\infty}$ for (\ref{eq:eqn14}) to
be a contraction map. We have
\begin{equation} \label{eq:eqn17}
\lVert\textbf{J}^{BR}\rVert_\infty=\max_{k \in
\mathcal{K}}\sum_{i=1}^K\lvert J_{ki}^{BR}\lvert \
\leq\frac{1}{2}+\frac{1}{2}\max_{k \in \mathcal{K}}\frac{K-1}{a_k}.
\end{equation}
Therefore, if $a_k>K-1, \forall k \in \mathcal{K}$, Algorithm 1 also
globally converges. However, it is easy to verify that it is a
special case of the sufficient condition given by (\ref{eq:eqn12}).
In addition, we can see from (\ref{eq:eqn12}) that, if the
accumulated ``aggressiveness" of the nodes in the entire networks
reaches a certain threshold, the global convergence property may not
hold. However, if all the nodes back off adequately by choosing
their algorithm parameters $\{a_k\}_{k=1}^K$ such that condition
(\ref{eq:eqn12}) is satisfied, Algorithm 1 globally converges.
\end{remark}

\begin{remark} \label{rm:rm4} Under the sufficient condition in (\ref{eq:eqn12}),
by substituting (\ref{eq:eqn8}) into (\ref{eq:eqn12}), the limiting
points lie in the set
\begin{equation} \label{eq:eqn18}
\Big\{ \textbf{p}^*=(p_1^*,\ldots,p_K^*)\Big| \sum_{i \in
\mathcal{K}\backslash\{k\}}\frac{p_i^*}{\prod_{l \in
\mathcal{K}\backslash\{i\}}(1-p_l^*)}<1, \forall k \in \mathcal{K}
\Big\}.
\end{equation}It is easy to check that this is a subset of $\{ \textbf{p}^*=(p_1^*,\ldots,p_K^*) | \sum_{k=1}^K
p_k^*<1\}$ for $K>2$, which verifies the intuition that the set that
Algorithm 1 globally converges to should be a subset of the set of
locally stable CE.
\end{remark}

\subsection{A Gradient Play Learning Algorithm}
The best-response based dynamics may lead to large fluctuations in
the entire network, which may not be desirable if we want to avoid
temporary system-wide instability. Therefore, in this subsection, we
propose an alternative learning algorithm inspired by the gradient
type dynamics, which has been well studied in the field of
evolutionary biology \cite{gradient_1}\cite{gradient_2}. For
example, in population genetics, the evolutionary dynamics resulting
for a particular mutant's invasion fitness, i.e. its growth rate,
are primarily governed by the fitness gradient. In other words, the
population has a small probability of moving its phenotype
\cite{Gene} in the direction in which fitness is increasing, and
this probability is proportional to the fitness gradient for
possible mutants. This model has also been used to model fluid flow
under a pressure gradient or the motion of organisms towards sites
of higher nutrient concentration \cite{gradient_3}.

Motivated by the gradient dynamics, we consider the gradient play
learning algorithm. At each iteration, each node updates its action
gradually in the ascent direction of its conjectured utility
function in (\ref{eq:eqn6}). Specifically, at stage $t$, node $k$
chooses its transmission probability according to
\begin{equation}
\label{eq:eqn21} p_k^t=\biggl[p_k^{t-1}+ \gamma_k\frac{\partial
u_k^t(\tilde{s}_k^t(p_k),p_k)}{\partial
p_k}\Biggr\rvert_{p_k=p_k^{t-1}}\biggr]_0^1,
\end{equation}
in which $[x]_a^b$ means $\max\{\min\{x,b\},a\}$. The engineering
interpretation of this updating procedure is that each node will
``mutate", i.e. update its transmission probability, along the
gradient direction of its conjectured utility function. As long as
the stepsize $\gamma_k$ is small enough, the entire network will
``evolve" smoothly and temporary system-wide instability will not
occur. This algorithm also resembles the technique of exponential
smoothing in statistics \cite{SM}. In the following, we assume that
all nodes use the same stepsize $\gamma_k = \gamma, \forall k\in
\mathcal {K}$ and $0<p_k^{t-1}<1$. If $\gamma$ is sufficiently
small, substituting the utility function (\ref{eq:eqn6}) into
(\ref{eq:eqn21}), we have
\begin{equation}
\label{eq:eqn22} p_k^t=p_k^{t-1}+ \gamma \Bigl\{\prod_{i \in
\mathcal{K}\backslash\{k\}}(1-p_i^{t-1})-a_kp_k^{t-1}\Bigr\}.
\end{equation}
The detailed description of the distributed gradient play learning
mechanism is summarized in Algorithm 2. As for Algorithm 1, we
investigate the stability and convergence of this gradient play
learning algorithm.
\begin{algorithm}
\caption{: A Distributed Gradient Play Learning Algorithm for Random
Access}
\begin{algorithmic}[1]
\State Initialize: $t=0$, stepsize $\gamma$, the transmission
probability $p_k^0\in[0,1]$, and the parameter $a_k>0$ in node $k$'s
belief function, $\forall k\in\mathcal{K}$. \Procedure {}{} \State
\emph{Locally} at each node $k$, iterate through $t$: \State Set $t
\gets t+1.$ \ForAll {$k\in\mathcal{K}$} \State At stage $t$,
$p_k^t\gets \biggl[p_k^{t-1}+ \gamma\Bigl\{\prod_{i \in
\mathcal{K}\backslash\{k\}}(1-p_i^{t-1})-a_kp_k^{t-1}\Bigr\}\biggr]_0^1$.
\EndFor \State Node $k$ decides if it will transmit data with a
 probability $p_k^t$ (or equivalently, maintain a window size of $CW_k^t=2/p_k^t-1$) for all the time slots during stage $t$. \EndProcedure
\end{algorithmic}
\end{algorithm}
\subsubsection{Local Stability}


First of all, the following theorem describes a stable CE set in $P$
for Algorithm 2.

\begin{theorem} \label{th:th7}
For any $\textbf{p}^*=(p_1^*,\ldots,p_K^*)\in P$, if
\begin{equation} \label{eq:eqnth7}
\sum_{k=1}^K p_k^*<1, \quad \textrm{or} \quad \sum_{i \in
\mathcal{K}\backslash\{k\}}\frac{p_k^*}{1-p_i^*}<1, \ \forall k \in
\mathcal{K},
\end{equation}and the stepsize $\gamma$ is sufficiently small, $\textbf{p}^*$ is a
stable CE for Algorithm 2.
\end{theorem}

\emph{Proof}: Consider the Jacobian matrix $\textbf{J}^{GP}$ of the
self-mapping function in (\ref{eq:eqn22}). We have
$J_{ik}^{GP}=\partial p_i^t/\partial p_k^{t-1}$. As discussed above,
for $\textbf{p}^*=(p_1^*,\ldots,p_K^*)\in P $ to be a fixed point of
the self-mapping function in (\ref{eq:eqn22}), $a_k$ must be set to
be $a_k^*=\prod_{i \in \mathcal{K}\backslash\{k\}}(1-p_i^*)/p_k^*$.
It follows that
\begin{equation} \label{eq:eqn23}
J_{ik}^{GP}\rvert_{\textbf{p}=\textbf{p}^*,\
\textbf{\textit{a}}=\textbf{\textit{a}}^*}=\left\{ \begin{array}{cl}
1-\gamma/p_k^* \cdot \prod_{l \in \mathcal{K}\backslash\{k\}}(1-p_l^*),& \text{if $i=k$},\\
-\gamma\prod_{l \in \mathcal{K}\backslash\{i,k\}}(1-p_l^*),&
\text{if $i\neq k$}.
\end{array} \right.
\end{equation}
$\textbf{p}^*$ is stable if and only if the eigenvalues
$\{\lambda_k\}_{k=1}^{K}$ of matrix $\textbf{J}^{GP}$ are all inside
the unit circle of the complex plane, i.e. $|\lambda_k|<1, \forall k
\in \mathcal{K}$. Recall that the spectral radius $\rho(\textbf{J})$
of a matrix $\textbf{J}$ is the maximal absolute value of the
eigenvalues\cite{Matrix_book}. Therefore, it is equivalent to prove
that $\rho(\textbf{J}^{GP})<1$.

To a vector $\pmb{w}=(w_1,\cdots,w_K) \in \mathcal{R}_+^K$ with
positive entries, we associate a \emph{weighted} $\ell_{\infty}$
\emph{norm}, defined as
\begin{equation} \label{eq:eqn24}
\|\pmb{x}\|_{\infty}^{\pmb{w}}=\max_{k \in
\mathcal{K}}\frac{|x_k|}{w_k}.
\end{equation}
The vector norm $\|\cdot\|_{\infty}^{\pmb{w}}$ induces a matrix
norm, defined by
\begin{equation} \label{eq:eqn25}
\|A\|_{\infty}^{\pmb{w}}=\max_{k \in
\mathcal{K}}\frac{1}{w_k}\sum_{i=1}^{K}|a_{ki}|w_i.
\end{equation}
According to Proposition A.20 in \cite{Bertsekas},
$\rho(\textbf{J}^{GP})\leq\|\textbf{J}^{GP}\|_{\infty}^{\pmb{w}}$.
Consider the vector $\pmb{w}=(w_1,\cdots,w_K)$ in which
$w_k=p_k^*(1-p_k^*)$. We have
\begin{equation} \label{eq:eqn26}
\begin{split}
\frac{1}{w_k}\sum_{i=1}^{K}|J_{ki}^{GP}|w_i&=1-\frac{\gamma \prod_{l
\in \mathcal{K}\backslash\{k\}}(1-p_l^*)}{p_k^*}+\sum_{i\in
\mathcal{K}\backslash\{k\}} \frac {\gamma p_i^* \prod_{l \in
\mathcal{K}\backslash\{k\}}(1-p_l^*)}{p_k^*(1-p_k^*)}\\
&=1-\frac{\gamma \prod_{l \in
\mathcal{K}\backslash\{k\}}(1-p_l^*)}{p_k^*}\Bigl[ 1- \sum_{i\in
\mathcal{K}\backslash\{k\}} \frac{p_i^*}{1-p_k^*}\Bigr].
\end{split}
\end{equation}

Therefore, if $\sum_{k \in \mathcal{K}}p_k^*<1, \ \forall k \in
\mathcal{K}$, there exists some $\beta>0$ such that
\begin{equation}
\frac{\prod_{l \in
\mathcal{K}\backslash\{k\}}(1-p_l^*)}{p_k^*}\Bigl[ 1- \sum_{i\in
\mathcal{K}\backslash\{k\}} \frac{p_i^*}{1-p_k}^*\Bigr]\geq \beta,
\quad \forall k\in \mathcal{K}.
\end{equation}
If the stepsize $\gamma$ satisfies $0<\gamma<1/\beta$, we have
\begin{equation}
\|\textbf{J}^{GP}\|_{\infty}^{\pmb{w}}=\max_{k \in
\mathcal{K}}\biggl\{1-\frac{\gamma \prod_{l \in
\mathcal{K}\backslash\{k\}}(1-p_l^*)}{p_k^*}\Bigl[ 1- \sum_{i\in
\mathcal{K}\backslash\{k\}} \frac{p_i^*}{1-p_k^*}\Bigr]\biggr\}\leq
1-\gamma\beta<1.
\end{equation}
Since
$\rho(\textbf{J}^{GP})\leq\|\textbf{J}^{GP}\|_{\infty}^{\pmb{w}}<1$,
all the eigenvalues of $\textbf{J}^{GP}$ must fall into the unit
circle $|\lambda|<1$. Therefore, $\textbf{p}^*$ is a stable CE.
Similarly, by choosing $\pmb{w}=[1,\cdots,1]$, we can show that, if
\begin{equation}
\sum_{i \in \mathcal{K}\backslash\{k\}}\frac{p_k^*}{1-p_i^*}<1, \
\forall k \in \mathcal{K},
\end{equation}and $\gamma$ is sufficiently small, $\textbf{p}^*$ is also stable. $\blacksquare$

\subsubsection{Global Convergence}
Similarly as in the previous subsection, we derive in the following
theorem a sufficient condition under which Algorithm 2 globally
converges.

\begin{theorem} \label{th:th6}
Regardless of any initial value chosen for $\{p_k^0\}_{k=1}^{K}$, if
the parameters $\{a_k\}_{k=1}^{K}$ in the belief functions
$\{\tilde{s}_k\}_{k=1}^{K}$ satisfy
\begin{equation} \label{eq:eqn32}
\sum_{i \in \mathcal{K}\backslash\{k\}}\frac{1}{a_i}<1, \forall k
\in \mathcal{K},
\end{equation}
and the stepsize $\gamma$ is sufficiently small, Algorithm 2
converges to a unique CE.
\end{theorem}

\emph{Proof}: For the self-mapping function in (\ref{eq:eqn22}), the
elements of its Jacobian matrix $\textbf{J}^{GP}$ satisfy
\begin{equation} \label{eq:eqn33}
J_{ik}^{GP}=\left\{ \begin{array}{cl}
1-\gamma a_k,& \text{if $i=k$},\\
-\gamma\prod_{l \in \mathcal{K}\backslash\{i,k\}}(1-p_l),& \text{if
$i\neq k$}.
\end{array} \right.\end{equation}

Consider the induced distance by weighted $\ell_{\infty}$ norm in
the Euclidean space. We have
\begin{equation} \label{eq:eqn34}
\lVert\textbf{p}^t-\hat{\textbf{p}}^t
\rVert_{\infty}^{\pmb{w}}\leq\lVert\textbf{J}^{GP}\rVert_{\infty}^{\pmb{w}}\cdot
\lVert\textbf{p}^{t-1}-\hat{\textbf{p}}^{t-1}
\rVert_{\infty}^{\pmb{w}}.
\end{equation}

Using $\pmb{w}=(1/a_1,\cdots,1/a_K)$ for (\ref{eq:eqn34}), we have
\begin{equation} \label{eq:eqn35}
\lVert\textbf{J}^{GP}\rVert_{\infty}^{\pmb{w}}=\max_{k \in
\mathcal{K}}\biggl\{ 1-\gamma a_k + \sum_{i\in
\mathcal{K}\backslash\{k\}}\prod_{l \in
\mathcal{K}\backslash\{i,k\}}\frac{\gamma a_k(1-p_l)}{a_i}\biggr\} \
\leq \max_{k \in \mathcal{K}}\biggl\{ 1-\gamma a_k \Bigl(1-
\sum_{i\in \mathcal{K}\backslash\{k\}}\frac{1}{a_i}\Bigr)\biggr\}.
\end{equation}
Therefore, if the condition in (\ref{eq:eqn32}) is satisfied, there
exists some $\beta>0$ such that
\begin{equation}
a_k \Bigl(1- \sum_{i\in
\mathcal{K}\backslash\{k\}}\frac{1}{a_i}\Bigr)\geq \beta, \quad
\forall k\in \mathcal{K}.
\end{equation}
If the stepsize $\gamma$ satisfies $0<\gamma<1/\beta$, we have
\begin{equation}
\|\textbf{J}^{GP}\|_{\infty}^{\pmb{w}}\leq\max_{k \in
\mathcal{K}}\biggl\{1-\gamma a_k \Bigl(1- \sum_{i\in
\mathcal{K}\backslash\{k\}}\frac{1}{a_i}\Bigr)\biggr\}\leq
1-\gamma\beta<1.
\end{equation}
Therefore, there exist a constant $q\in[0,1)$ and a positive
$\epsilon$, such that
$q=\lVert\textbf{J}^{GP}\rVert_{\infty}^{\pmb{w}}=1-\epsilon<1$ and
$\lVert\textbf{p}^t-\hat{\textbf{p}}^t \rVert_{\infty}^{\pmb{w}}\leq
q \lVert\textbf{p}^{t-1}-\hat{\textbf{p}}^{t-1}
\rVert_{\infty}^{\pmb{w}}$. From the contraction mapping theorem
\cite{Contraction_mapping}, the self-mapping function in
(\ref{eq:eqn22}) has a unique fixed point and the sequence
$\{\textbf{p}^t\}_{t=0}^{+\infty}$ converges to the unique fixed
point. \ $\blacksquare$

\begin{remark}
Compare Theorem \ref{th:th7} and \ref{th:th6} with Theorem
\ref{th:th2} and \ref{th:th4}. We can see that, given the same
target operating point $\textbf{p}$ or parameters $\{a_k\}_{k=1}^K$,
Algorithm 2 exhibits similar properties in terms of local stability
and global convergence, provided that its stepsize $\gamma$ is
sufficiently small. In other words, the limiting behavior of these
two distinct bio-inspired dynamic mechanisms are similar. However,
we need to consider some design trade-off for both algorithms and
choose the desired learning algorithm based on the specific system
requirements about the speed of convergence and the performance
fluctuation. Generally speaking, the best response learning
algorithm converges fast, but it may cause temporary large
fluctuations during the convergence process, which is not desirable
for transporting constant-bit-rate applications. On the other hand,
the gradient play learning algorithm with small stepsize will evolve
smoothly at the cost of sacrificing its convergence rate.
\end{remark}

\subsection{Alternative Interpretations of the Conjecture-based Learning Algorithms}
\begin{figure}
\centering
\includegraphics[width=0.7\textwidth]{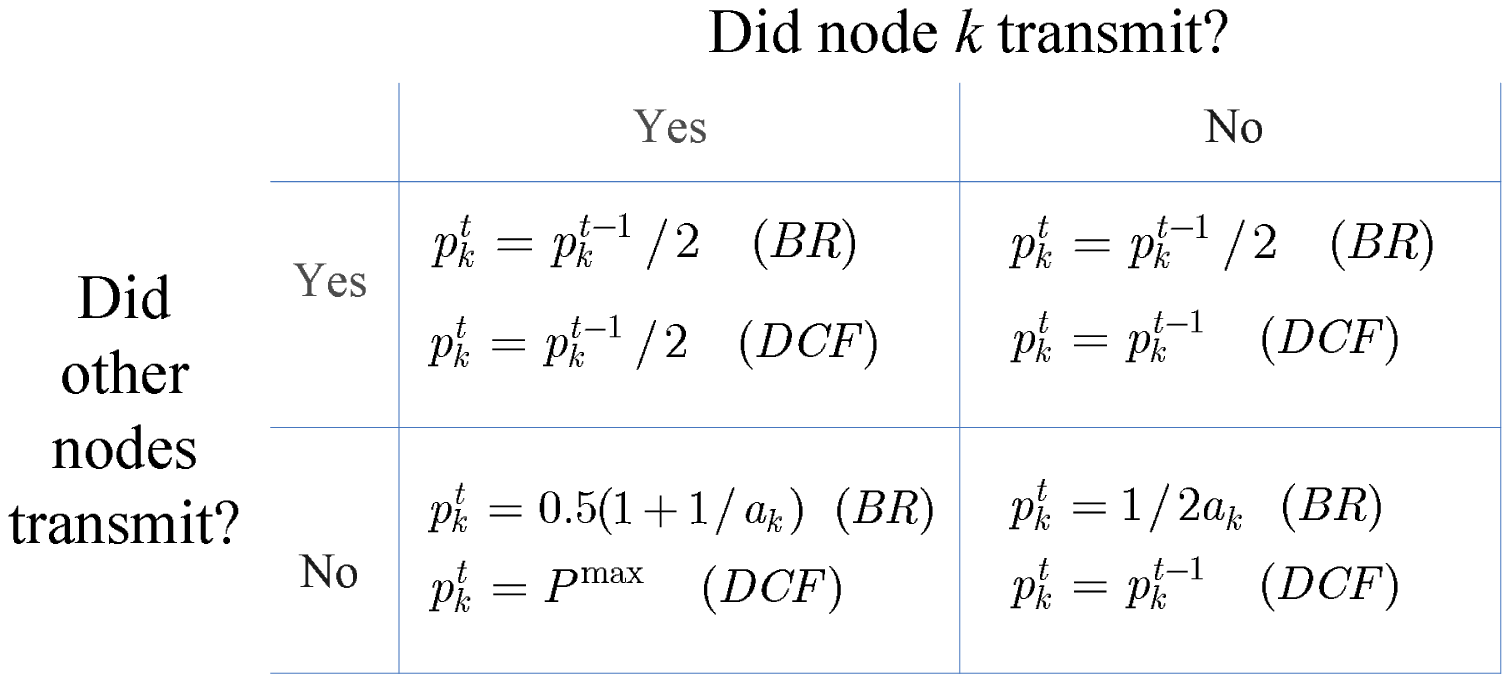}
\caption{Comparison between the best response learning and the IEEE
802.11 DCF ($P^{\max}$ is specified in the DCF protocol).}
\label{fg:DCFvsBR}
\end{figure}

In this section, we re-interpret the proposed algorithms using the
the backoff mechanism model in which the transmission probabilities
change from time slot to time slot \cite{Mobicom_ref}, which helps
us to understand the key difference between the proposed algorithms
and 802.11 DCF. The superscript $t$ in this subsection represents
the numbering of the time slots. We define $T_k^t$ and $T_{-k}^t$ as
the events that node $k$ transmits data at time slot $t$ and any
node in $\mathcal{K}\backslash\{k\}$ transmits data at time slot
$t$, respectively. If $a_k>1$, the RHS of (\ref{eq:eqn7}) equals to
$\frac{p_k^{t-1}}{2}+\frac {\prod_{i \in
\mathcal{K}\backslash\{k\}}(1-p_i^{t-1})}{2a_k}$, and the best
response update function in (\ref{eq:eqn7}) can be rewritten as
\begin{equation}\label{eq:reform}
p_k^t=\frac{1}{2}\textrm{E}\{p_k^{t-1}\textbf{1}_{\{T_{-k}^{t-1}=1\}}|\textbf{p}^{t-1}\}+\frac
{1}{2a_k}\textrm{E}\{\textbf{1}_{\{T_{-k}^{t-1}=0\}}\textbf{1}_{\{T_{k}^{t-1}=0\}}|\textbf{p}^{t-1}\}+\frac
{1}{2}(1+\frac{1}{a_k})\textrm{E}\{\textbf{1}_{\{T_{-k}^{t-1}=0\}}\textbf{1}_{\{T_{k}^{t-1}=1\}}|\textbf{p}^{t-1}\},
\end{equation}
where $\textbf{1}_a$ is an indicator function of event $a$ taking
place, $\textrm{E}\{a|b\}$ is the expected value of $a$ given $b$,
$\textrm{E}\{\textbf{1}_{\{T_{k}^{t-1}=1\}}|\textbf{p}^{t-1}\}=p_k^{t-1}$,
and
$\textrm{E}\{\textbf{1}_{\{T_{-k}^{t-1}=1\}}|\textbf{p}^{t-1}\}=1-\prod_{i
\in \mathcal{K}\backslash\{k\}}(1-p_i^{t-1})$. According to
(\ref{eq:reform}), we can provide an alternative interpretation of
the best-response update algorithm as follows. Consider the
following update algorithm. At each time slot, if node $k$ observes
that any other node attempts to transmit, i.e. it senses a busy
channel, it reduces its transmission probability by a factor $1/2$.
If no transmission attempt is made by any node in the system, node
$k$ sets its transmission probability to be $1/2a_k$. Otherwise, if
node $k$ makes a successful transmission, it will transmit with
probability $0.5(1+1/a_k)$ in the next time slot. We can see that
equation (\ref{eq:reform}) characterizes the expected trajectory of
this alternative update mechanism. Fig. \ref{fg:DCFvsBR} compares
this new interpretation with the IEEE 802.11 DCF \cite{Rvs_Num}. We
can see that, node $k$ behaves similarly in the best response
algorithm and the IEEE 802.11 DCF if it made a transmission attempt
in the previous time slot, and the fundamental difference between
these two protocols is how node $k$ updates its action given that it
did not transmit in the previous time slot. In DCF, $p_k^t$ is kept
the same as $p_k^{t-1}$. However, as we can see from
(\ref{eq:reform}), the best response algorithm either performs
back-off if the channel is busy or sets $p_k^t$ to be $1/2a_k$ if
the channel is free. This can also be intuitively interpreted from a
biological perspective: if the channel is busy, meaning that other
competitors are accessing the resource (transmission opportunity),
the node can avoid a confrontation by becoming less aggressive (i.e.
reducing its transmission probability); if on the other hand, the
system is idle and the resource is wasted, the node will consume the
resource by increasing its transmission probability to $1/2a_k$.

\begin{remark} Both Equation (\ref{eq:eqn8}) and (\ref{eq:reform})
intuitively explain the meaning of the algorithm parameters
$\{a_k\}_{k=1}^K$. Note that the numerator of (\ref{eq:eqn8}),
$\prod_{i \in \mathcal{K}\backslash\{k\}}(1-p_i^*)$, represents the
probability that transmitter $k$ experiences a contention-free
environment at $\textbf{p}^*$. The value of $1/a_k$, i.e. the ratio
between node $k$'s transmission probability $p_k$ and its
contention-free probability, indicates the ``aggressiveness" of this
particular node at equilibrium. In addition, according to
(\ref{eq:reform}), the transmission probability $1/2a_k$ also
reflects node $k$'s ``aggressiveness" in selecting its transmission
probability after it sensed a free channel. It is straightforward to
see the selection of $\{a_k\}_{k=1}^K$ introduces some trade-off
between the stability and throughput of the networks. First of all,
large values of $\{a_k\}_{k=1}^K$ refrain nodes from transmitting at
a higher channel access probability, and hence, it stabilizes the
system at the cost of reducing the throughput. On the other hand,
lowering $\{a_k\}_{k=1}^K$ increases the nodes' transmission
probability, which may improve the throughput performance. However,
it can cause the conditions in (\ref{eq:eqn9}) and (\ref{eq:eqnth7})
to fail and the system becomes unstable. Therefore, the problems
which we will investigate in the next subsection are which part of
the throughput region can be achieved with stable CE and how the
nodes can adaptively update their $\{a_k\}_{k=1}^K$ such that the
system can attain efficient and stable operating points.
\end{remark}

Before proceeding to the next subsection, similarly as for the best
response algorithm, we present an reinterpretation of the gradient
play. Equation (\ref{eq:eqn22}) can be rewritten as
\begin{equation}\label{eq:reform2}
p_k^t=(1-\gamma
a_k)p_k^{t-1}\textrm{E}\{\textbf{1}_{\{T_{-k}^{t-1}=1\}}|\textbf{p}^{t-1}\}+[p_k^{t-1}+\gamma
(1-a_kp_k^{t-1})]\textrm{E}\{\textbf{1}_{\{T_{-k}^{t-1}=0\}}|\textbf{p}^{t-1}\}.
\end{equation}
If $a_k>1$, the interpretation of (\ref{eq:reform2}) is that at each
time slot, if node $k$ senses a busy channel, it reduces its
transmission probability by a factor $1-\gamma a_k$, otherwise it
increases its transmission probability by an amount $\gamma
(1-a_kp_k^{t-1})$. We can see that, this interpretation of the
gradient play learning resembles the well-known AIMD (Additive
Increase Multiplicative Decrease) control algorithm, which has been
widely applied in the context of congestion avoidance in computer
networks due to its superior performance in terms of convergence and
efficiency \cite{AIMD}.

\subsection{Stability of the Throughput Region}

The results in the previous subsections describe the values of
$\{p_k\}_{k=1}^{K}$ and $\{a_k\}_{k=1}^{K}$ for which local
stability and global convergence can be guaranteed in both Algorithm
1 and 2. This subsection directly investigates for both algorithms
the stability of achievable operating points in the throughput
region $\mathscr{T}$.

\begin{lemma} \label{lm:lm1}
The Pareto boundary of the throughput region $\mathscr{T}$ is the
set of all points $\pmb{\tau}=(\tau_1,\ldots,\tau_K)$ such that
$\tau_k=p_k \prod_{i\in \mathcal{K}\backslash\{k\}}(1-p_i)$ where
$\textbf{p}=(p_1,\ldots,p_K)$ is a vector satisfying
$\textbf{p}\geqslant\textbf{0}$ and $\sum_{k\in \mathcal {K}}p_k=1$;
and each such $\pmb{\tau}$ is determined by a unique such
$\textbf{p}$.
\end{lemma}

\emph{Proof}: See Theorem 1 in \cite{IT_throughput}. \
$\blacksquare$

\begin{theorem} \label{th:th3}
Regardless of the number of nodes in the network, for any
Pareto-inefficient operating point $\pmb{\tau}^*$ in the throughput
region $\mathscr{T}$, there always exists a belief configuration
$\{a_k\}_{k=1}^{K}$ stabilizing Algorithm 1 and 2, and achieve the
throughput $\pmb{\tau}^*$. If $K>2$, any Pareto-optimal operating
point $\{p_k^*\}_{k=1}^K$ in $\mathscr{T}$ that satisfies $p_k^*>0 ,
\forall k\in \mathcal{K}$ is a stable CE for Algorithm 1 and 2.
\end{theorem}

\emph{Proof}: From Theorem \ref{th:th2}, we know that $\sum_{k \in
\mathcal {K}}p_k<1$ is sufficient to guarantee that the
corresponding CE is stable. Therefore, it is equivalent to check
that any Pareto-inefficient operating point $\pmb{\tau}^*$ can be
achieved with a joint transmission probability $\textbf{p}^*\in P$
satisfying $\sum_{k \in \mathcal {K}} p_k^*<1$.

Define the throughput region
\begin{equation} \label{eq:eqn19}
\mathscr{T}(t)=\{(u_1(\textbf{p}),\ldots,\\
u_K(\textbf{p}))| \ \exists \ \textbf{p}\in P, \sum_{k \in \mathcal
{K}}p_k\leq t\},
\end{equation} in which an additional constraint $\sum_{k \in \mathcal
{K}}p_k\leq t$ is imposed. We denote the Pareto boundary of
$\mathscr{T}(t)$ as
\begin{equation} \label{eq:eqn20}
\partial\mathscr{T}(t)=\{\pmb{\tau}| \ \nexists \ \pmb{\tau}'\in \mathscr{T}(t) \ \textrm{such
that} \ \tau'_k\geq\tau_k, \ \forall k \in \mathcal {K}\
\textrm{and} \ \tau'_k>\tau_k, \ \exists k \in \mathcal {K}\}.
\end{equation}
Following the proof of Lemma \ref{lm:lm1}, we can draw a similar
conclusion: all the points on $\partial\mathscr{T}(t)$ satisfy
$\sum_{k\in \mathcal {K}}p_k=t$. By Lemma \ref{lm:lm1},
$\partial\mathscr{T}(1)$ corresponds to the Pareto boundary of
$\mathscr{T}$. Note that $\partial\mathscr{T}(0)=\textbf{0}$. In
other words, varying $t$ from 1 to 0 will cause
$\partial\mathscr{T}(t)$ to continuously shrink from the Pareto
boundary of the throughput region $\mathscr{T}$ to the origin
$\textbf{0}$. Therefore, for any Pareto inefficient point
$\pmb{\tau}^*\in\mathscr{T}$, there exists $0\leq t'<1$ such that
$\pmb{\tau}^*$ lie on $\partial\mathscr{T}(t')$, i.e. $\pmb{\tau}^*$
can be achieved with an action profile $\textbf{p}^*$ satisfying
$\sum_{k\in \mathcal {K}}p_k^*=t<1$.

To prove the Pareto boundary are stable CE when $K>2$, we need to
show that the eigenvalues $\{\xi_n^{BR}\}_{n=1}^{N}$ of the Jacobian
matrices $\textbf{J}^{BR}$ and $\textbf{J}^{GP}$ are all inside the
unit circle of the complex plane \cite{Contraction_mapping}, i.e.
$|\xi_n|<1, \forall n \in \mathcal{N}$. Take the best response
dynamics for example. To determine the eigenvalues of
$\textbf{J}^{BR}$, we have
\begin{displaymath}
\begin{split}\det(\xi I-\textbf{J}^{BR})&=\left|
                              \begin{array}{cccc}
                                \xi-\frac{1}{2} & \frac {p_1}{2(1-p_2)} & \ldots & \frac {p_1}{2(1-p_K)} \\
                                \frac {p_2}{2(1-p_1)} & \xi-\frac{1}{2} & \ldots & \frac {p_2}{2(1-p_K)} \\
                                \vdots & \vdots & \ddots & \vdots \\
                                \frac {p_K}{2(1-p_1)} & \frac {p_K}{2(1-p_2)} & \ldots & \xi-\frac{1}{2} \\
                              \end{array}
                            \right| \\
&=\left|
                              \begin{array}{cccc}
                                \xi-\frac{1}{2} & \frac {p_1}{2(1-p_2)} & \ldots & \frac {p_1}{2(1-p_K)} \\
                                \frac {p_2}{2(1-p_1)}-\frac {p_2}{p_1}\bigl(\xi-\frac{1}{2}\bigr) & \xi-\frac{1}{2}-\frac {p_2}{2(1-p_2)} & \ldots & 0 \\
                                \vdots & \vdots & \ddots & \vdots \\
                                \frac {p_K}{2(1-p_1)}-\frac {p_K}{p_1}\bigl(\xi-\frac{1}{2}\bigr) & 0 & \ldots & \xi-\frac{1}{2}-\frac {p_K}{2(1-p_K)} \\
                              \end{array}
                            \right| \\
&=\left|
                              \begin{array}{cccc}
                                \bigl(\xi-\frac{1}{2}-\frac {p_1}{2(1-p_1)}\bigr)\cdot\bigl[1+\sum_{k=1}^K\frac{\frac {p_k}{2(1-p_k)}}{\xi-\frac{1}{2}-\frac {p_k}{2(1-p_k)}}\bigr] & 0 & \ldots & 0 \\
                                \frac {p_2}{2(1-p_1)}-\frac {p_2}{p_1}\bigl(\xi-\frac{1}{2}\bigr) & \xi-\frac{1}{2}-\frac {p_2}{2(1-p_2)} & \ldots & 0 \\
                                \vdots & \vdots & \ddots & \vdots \\
                                \frac {p_K}{2(1-p_1)}-\frac {p_K}{p_1}\bigl(\xi-\frac{1}{2}\bigr) & 0 & \ldots & \xi-\frac{1}{2}-\frac {p_K}{2(1-p_K)} \\
                              \end{array}
                            \right|. \end{split}\end{displaymath}
Therefore, we can see that, the eigenvalues of $\textbf{J}^{BR}$ are
the roots of
\begin{equation}\label{eq:eqn201}
\bigl[1+\sum_{k=1}^K\frac{\frac
{p_k}{2(1-p_k)}}{\xi-\frac{1}{2}-\frac
{p_k}{2(1-p_k)}}\bigr]\cdot\prod_{k=1}^K\bigl(\xi-\frac{1}{2}-\frac
{p_k}{2(1-p_k)}\bigr)=0.
\end{equation}

Denote $f(\xi)=\sum_{k=1}^K\frac{\frac
{p_k}{2(1-p_k)}}{\xi-\frac{1}{2}-\frac {p_k}{2(1-p_k)}}$. First, we
assume that $p_i\neq p_j, \forall i,j$. Without loss of generality,
consider $p_1<p_2<\cdots<p_K$. In this case, the eigenvalues of
$\textbf{J}^{BR}$ are the roots of $f(\xi)=-1$. Note that $f(\xi)$
is a continuous function and it strictly decreases in
$(-\infty,\frac{1}{2}+\frac {p_1}{2(1-p_1)})$, $(\frac{1}{2}+\frac
{p_1}{2(1-p_1)},\frac{1}{2}+\frac {p_2}{2(1-p_2)})$, $\cdots$,
$(\frac{1}{2}+\frac {p_{K-1}}{2(1-p_{K-1})},\frac{1}{2}+\frac
{p_K}{2(1-p_K)})$, and $(\frac {p_K}{2(1-p_K)},+\infty)$. We also
have
$\lim_{\xi\rightarrow(\frac{1}{2}+\frac{p_k}{2(1-p_k)})^-}f(\xi)=-\infty$,
$\lim_{\xi\rightarrow(\frac{1}{2}+\frac{p_k}{2(1-p_k)})^+}f(\xi)=+\infty,
n=1,2,\cdots,K$, and
$\lim_{\xi\rightarrow-\infty}f(\xi)=\lim_{\xi\rightarrow+\infty}f(\xi)=0$.
Therefore, the roots of $f(\xi)=-1$ lie in
$(-\infty,\frac{1}{2}+\frac {p_1}{2(1-p_1)})$, $(\frac{1}{2}+\frac
{p_1}{2(1-p_1)},\frac{1}{2}+\frac {p_2}{2(1-p_2)})$, $\cdots$,
$(\frac{1}{2}+\frac {p_{K-1}}{2(1-p_{K-1})},\frac{1}{2}+\frac
{p_K}{2(1-p_K)})$ respectively.

For the operating points on the Pareto boundary, we have
$\sum_{k=1}^K p_k=1$. It is easy to verify that $f(0)=-1$, i.e.
$\xi_1=0$. Therefore,
\begin{equation}
\rho(\textbf{J}^{BR})=\max_k|\xi_k|=\xi_K\in (\frac{1}{2}+\frac
{p_{K-1}}{2(1-p_{K-1})},\frac{1}{2}+\frac {p_K}{2(1-p_K)}).
\end{equation}

To see $\xi_K<1$ for $0\leq p_1<p_2<\cdots<p_K$ and $\sum_{k=1}^K
p_k=1$. We differentiate two cases:

1) If $p_K\leq 0.5$, we have $\xi_K<\frac{1}{2}+\frac
{p_K}{2(1-p_K)}\leq 1$;

2) If $p_K>0.5$, we have $\frac{1}{2}+\frac
{p_{K-1}}{2(1-p_{K-1})}<1$ and $\frac{1}{2}+\frac
{p_K}{2(1-p_K)}>1$. Since $f(\xi)$ strictly decreases in
$(\frac{1}{2}+\frac {p_{K-1}}{2(1-p_{K-1})},\frac{1}{2}+\frac
{p_K}{2(1-p_K)})$, we have $\rho(\textbf{J}^{BR})<1$ if and only if
$f(1)<-1$. In fact,
\begin{equation}
f(1)-(-1)=\sum_{k=1}^K\frac{p_k}{1-2p_k}+1=\sum_{k=1}^{K-1}p_k\bigl(\frac{1}{1-2p_k}-\frac{1}{1-2\sum_{m=1}^{K-1}p_m}\bigr)<0.
\end{equation}
The inequality holds because
$\frac{1}{1-2p_k}<\frac{1}{1-2\sum_{m=1}^{K-1}p_m}$ for
$k=1,2,\ldots,K-1$ when $p_K>0.5$, $0\leq p_1<p_2<\cdots<p_K$, and
$\sum_{k=1}^K p_k=1$.

Second, we consider the cases in which there exists $p_i=p_j$ for
certain $i,j$. Suppose that $\{p_k\}_{k=1}^{K}$ take $M$ discrete
values $\kappa_1,\cdots,\kappa_M$ and the number of
$\{p_k\}_{k=1}^{K}$ that equal to $\kappa_m$ is $n_m$. In this case,
Equation (\ref{eq:eqn201}) is reduced to
\begin{equation}
\bigl[1+\sum_{k=1}^K\frac{\frac
{p_k}{2(1-p_k)}}{\xi-\frac{1}{2}-\frac
{p_k}{2(1-p_k)}}\bigr]\cdot\prod_{m=1}^M\bigl(\xi-\frac{1}{2}-\frac
{\kappa_m}{2(1-\kappa_m)}\bigr)^{n_m}=0.
\end{equation}
Hence, equation $f(\xi)=-1$ has $K+M-\sum_{k=1}^Mn_m$ roots in
total, and $\xi=\frac{1}{2}+\frac {\kappa_m}{2(1-\kappa_m)}$ is a
root of multiplicity $n_m-1$, $\forall m$. All these roots are the
eigenvalues of matrix $\textbf{J}^{BR}$. Similarly, the remaining
roots of $f(\xi)=-1$ lie in $(-\infty,\frac{1}{2}+\frac
{\kappa_1}{2(1-\kappa_1)})$, $(\frac{1}{2}+\frac
{\kappa_1}{2(1-\kappa_1)},\frac{1}{2}+\frac
{\kappa_2}{2(1-\kappa_2)})$, $\cdots$, $(\frac{1}{2}+\frac
{\kappa_{M-1}}{2(1-\kappa_{M-1})},\frac{1}{2}+\frac
{\kappa_M}{2(1-\kappa_M)})$. If $K>2$, $\sum_{k=1}^K p_k=1$ is still
sufficient to guarantee that $f(1)<-1$. Therefore, $|\xi_k^{BR}|<1,
\forall k \in \mathcal{K}$. $\blacksquare$

\begin{figure}
\centering
\includegraphics[width=0.5\textwidth]{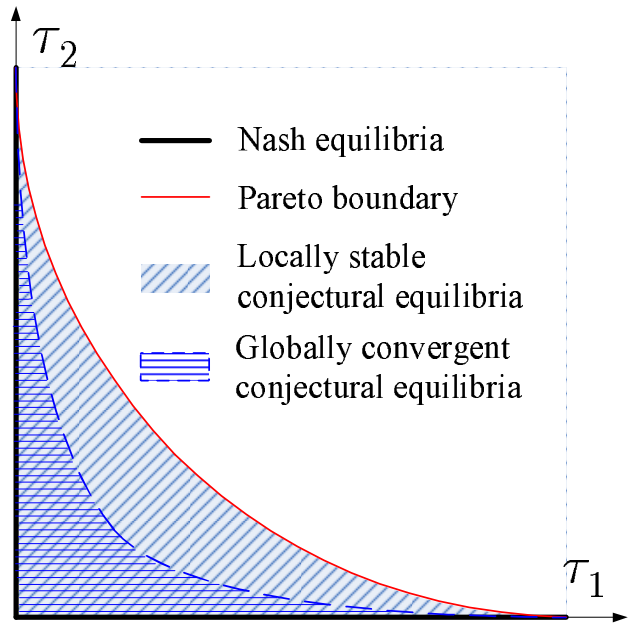}
\caption{Comparison among different solution concepts.}
\label{fg:region}
\end{figure}

Fig. \ref{fg:region} compares the throughput performance among
various game-theoretic solution concepts, including Nash equilibria,
Pareto frontier, locally stable conjectural equilibria, and globally
convergent conjectural equilibria,  in random access games. As
proven in Theorem \ref{th:th3}, Fig. \ref{fg:region} shows that, the
entire space spanning between the Nash equilibria and Pareto
frontier essentially consists of stable conjectural equilibria. In
addition, as discussed in Remark \ref{rm:rm4}, the set of globally
convergent CE is a subset of the stable CE set.

In practice, it is more important to construct algorithmic
mechanisms to attain the desirable CE that operate stably and
closely to the Pareto boundary. To this end, we develop an iterative
algorithm and summarize it as Algorithm 3. Specifically, this
algorithm has an inner loop and an outer loop. The inner loop adopts
either Algorithm 1 or 2 to achieve convergence for fixed
$\{a_k\}_{k=1}^K$. This algorithm initializes $a_k>|\mathcal{K}|$
such that it initially globally converges. After converging to a
stable CE, the outer loop adaptively adjusts $\{a_k\}_{k=1}^K$ until
desired efficiency is attained. The outer loop updates
$\{a_k\}_{k=1}^K$ in the multiplicative manner due to two reasons.
First, reducing $\{a_k\}_{k=1}^K$ individually increases
$\{p_k^t\}_{k=1}^K$ and $\sum_{k=1}^Kp_k$ and hence, moves the
operating point towards the Pareto boundary. Second, multiplying
$\{a_k\}_{k=1}^K$ by the same discount factor can maintain weighted
fairness among different nodes. Both reasons will be analytically
explained in the Section IV. It is also worth mentioning that
individual nodes can measure the Pareto efficiency in a fully
distributed manner during the outer loop iteration. For example,
individual nodes can estimate the other nodes' transmission
probabilities $\{p_k^t\}_{k=1}^K$ based on its local observation and
figure out whether the current operating point is close to the
Pareto boundary by calculating $\sum_{k\in\mathcal{K}}p_k^t$
\cite{Num_nomsg}. When the network size grows bigger, individually
estimating different nodes' transmission probabilities becomes
challenging. An alternative solution is that individual nodes can
instead monitor their common observation of the aggregate throughput
$\sum_{k\in\mathcal{K}}u_k^t$ and terminate the update of
$\{a_k\}_{k=1}^K$ once the aggregate throughput starts to decrease.
Next, we discuss several implementation issues regarding Algorithm
3. First, it is not necessary that all the nodes update their
parameters $\{a_k\}_{k=1}^K$ synchronously. However, these nodes
need to maintain the same update frequency, e.g. each node will
update its parameter after a certain number of timeslots or seconds.
As long as $\delta$ is small, the performance gap between the actual
CE and the intended one will not be large. Moreover, in order to
guarantee fairness, the new incoming nodes need to know the
real-time parameters of the old nodes in the same traffic class.
This initialization only needs to be done once, when the new nodes
enter the cell by tracking the evolution of the transmission
probabilities of the nodes in the same traffic class.

\begin{algorithm}
\caption{: Adaptive Distributed Learning Algorithm for Random
Access}
\begin{algorithmic}[1]
\State Initialize: stepsize $\gamma$ and $\delta$, the transmission
probability $p_k^0\in[0,1]$, and the parameter $a_k>|\mathcal{K}|$
in node $k$'s belief function, $\forall k\in\mathcal{K}$. \Procedure
{}{} \State \textbf{outer loop:} For each node $k$, $a_k\leftarrow
a_k(1-\delta)$. \State \quad \textbf{inner loop:} \emph{Locally} at
each node $k$, use Algorithm 1 or 2 to update $p_k^t$. \State \quad
 \textbf{until} it converges. \State \textbf{until} the aggregate throughput is maximized or $\sum_{k\in\mathcal{K}}p_k^t\approx1$. \EndProcedure
\end{algorithmic}
\end{algorithm}

\section{Extensions to Heterogeneous Networks and Ad-hoc Networks}
In this section, we first investigate how users with different
qualify-of-service requirements should initialize their belief
functions and interact in the heterogeneous network setting and show
that the conjecture-based approaches approximately achieve the
weighted fairness. Furthermore, we discuss how the single-cell
solution can be extended to the general ad-hoc network scenario,
where only the devices within a certain neighborhood range will
impact each other's throughput.

\subsection{Equilibrium Selection for Heterogeneous Networks}

Consider a network with $N>1$ different classes of nodes. Let
$\phi_n$ denote the parameter that class-$n$ nodes choose for their
conjectured utility functions (i.e. the parameter $a_k$ if node $k$
belongs to class-$n$) and $\mathcal{F}_n$ denote the set of nodes
that set their algorithm parameters to be $\phi_n$, $1\leq n\leq N$.
At equilibrium, the transmission probabilities of the same class of
nodes are equal, denoted as $\tilde{p}_n$. Before we proceed, we
first define the weighted fairness for the random access game
\cite{fairness}. For each traffic class $n$, we associate with a
positive weight $\chi_n$. Then the weighted fairness intended for
the random access game satisfy
\begin{equation}\label{eq:wf1}
\forall i,j \in \{1,2,\cdots,N\}, \forall s \in \mathcal{F}_i,
\forall s' \in \mathcal{F}_j,
\frac{\textrm{E}\{\textbf{1}_{\{T_{-s}=0\}}\textbf{1}_{\{T_{s}=1\}}\}}{\chi_i}=\frac{\textrm{E}\{\textbf{1}_{\{T_{-s'}=0\}}\textbf{1}_{\{T_{s'}=1\}}\}}{\chi_j},
\end{equation}
which means that the probability of an successful transmission
attempt for traffic class $n$ is proportional to its weight
$\chi_n$. By simple manipulation, we have the equivalent form for
equation (\ref{eq:wf1}) \cite{fairness}:
\begin{equation}\label{eq:wf2}
\forall i,j \in \{1,2,\cdots,N\},
\frac{p_i^{WF}}{(1-p_i^{WF})\chi_i}=\frac{p_j^{WF}}{(1-p_j^{WF})\chi_j}.
\end{equation}

Recall that Theorem \ref{th:th1} showed how to choose
$\{a_k\}_{k=1}^K$ given a desired operating point
$\{p_k^*\}_{k=1}^K$ such that it is a CE. The following theorem
indicates the quantitative relationship between the chosen algorithm
parameters $\{\phi_n\}_{n=1}^N$, the sizes of different classes
$\{\mathcal{F}_n\}_{n=1}^N$, and the resulting steady-state
transmission probabilities $\{\tilde{p}_n\}_{n=1}^N$. More
importantly, it also shows that if the network size is large, the
conjecture-based algorithms approximately achieve weighted fairness.
\begin{theorem}\label{th:th9}
Suppose that $\phi_n\geq2,\ \forall 1\leq n\leq N$. The achieved
steady-state transmission probabilities $\{\tilde{p}_n\}_{n=1}^N$
are given by
\begin{equation}\label{eq:eqn38}
\tilde{p}_n=\frac{1}{2}\bigg(1-\sqrt{1-\frac{4\varrho}{\phi_n}}\bigg),
\end{equation}
where $\varrho$ satisfies
\begin{equation}\label{eq:eqn39}
\varrho=\frac{1}{2^K}\prod_{n=1}^N\bigg[\bigg(1+\sqrt{1-\frac{4\varrho}{\phi_n}}\bigg)^{|\mathcal{F}_n|}\bigg].
\end{equation}
\end{theorem}

\emph{Proof}: As shown in Theorem \ref{th:th1}, $
a_k^*p_k^*=\prod_{i \in \mathcal{K}\backslash\{k\}}(1-p_i^*)$.
Denote $\varrho=\prod_{i \in \mathcal{K}}(1-\tilde{p}_i)$.
Therefore, we obtain
\begin{equation}\label{eq:eqn40}
\phi_n\tilde{p}_n(1-\tilde{p}_n)=\varrho,\ \forall\ 1\leq n\leq N.
\end{equation}
Since $\phi_n>2$, we have $\tilde{p}_n<0.5$. Such a root of the
quadratic equation in (\ref{eq:eqn40}) is given in (\ref{eq:eqn38}).
Note that $\varrho=\prod_{i \in \mathcal{K}}(1-\tilde{p}_i)$.
Substituting (\ref{eq:eqn38}) into this equality, we get
(\ref{eq:eqn39}).

We can verify that a unique $\varrho$ satisfying the equality in
(\ref{eq:eqn39}) exists if $\phi_n\geq2,\ \forall 1\leq n\leq N$.
This is because the RHS of (\ref{eq:eqn39}) is feasible for
$\varrho\leq \textrm{min}_{1\leq n\leq N}\{\phi_n\}/4$ and it is a
strictly decreasing function in $\varrho$. Meanwhile, the LHS of
(\ref{eq:eqn39}) is strictly increasing on $\varrho \in [0,
\textrm{min}_{1\leq n\leq N}\{\phi_n\}/4]$. Note that when
$\varrho=\textrm{min}_{1\leq n\leq N}\{\phi_n\}/4$,
\begin{equation}
\textrm{LHS of (\ref{eq:eqn39})}=\min_{1\leq n\leq
N}\frac{\phi_n}{4}\geq \frac{1}{2}\geq \textrm{RHS of
(\ref{eq:eqn39})}.
\end{equation}
if $\phi_n\geq2$. Therefore, a unique $\varrho \in [0,
\textrm{min}_{1\leq n\leq N}\{\phi_n\}/4]$ satisfies
(\ref{eq:eqn39}) exists.\ $\blacksquare$

\begin{remark} There are several intuitions and observations that we
can obtain from Theorem \ref{th:th9}. First, the multiplicative
decreasing update in Algorithm 3 aims to move the operating points
towards Pareto boundary. A quantitative approximation between the
steady-state transmission probability $\tilde{p}_n$ and the
algorithm parameter $\phi_n$ of each traffic class can be derived if
a large number of nodes coexist. Since $\varrho \rightarrow0$ when
$|\mathcal{F}_n|$ is large, using the Taylor expansion,
$\tilde{p}_n$ can be approximated as $\varrho/\phi_n$, i.e. the
steady-state transmission probability $\tilde{p}_n$ decays as the
inverse first power of parameter $\phi_n$ that indicates the
``aggressiveness" of traffic class $n$. Finally, we also observe
from (\ref{eq:eqn40}) that, if $|\mathcal{F}_n|$ is large,
$\tilde{p}_n \rightarrow 0$ and $1-\tilde{p}_n\approx 1$. Therefore,
\begin{equation}\label{eq:wf3}
\forall i,j \in \{1,2,\cdots,N\},\phi_i
\tilde{p}_i(1-\tilde{p}_i)=\phi_j \tilde{p}_j(1-\tilde{p}_j)
\Rightarrow \frac{\phi_i\tilde{p}_i}{1-\tilde{p}_i} \approx
\frac{\phi_j \tilde{p}_j}{1-\tilde{p}_j}.
\end{equation}
Equation (\ref{eq:wf3}) indicates that Algorithm 1 and 2
approximately achieve weighted fairness given in (\ref{eq:wf2}) with
weight $\chi_n=1/\phi_n$. Moreover, it is worth mentioning that the
weighted fairness is purely an implicit by-product of the
conjecture-based approach and it can be sustained with stability.
Therefore, Algorithm 3 chooses to multiply $\{a_k\}_{k=1}^K$ by the
same discount factor $1-\delta$ such that the weighted fairness can
be maintained.
\end{remark}
\subsection{Extension to Ad-hoc Networks}

\begin{figure}
\centering
\includegraphics[width=0.6\textwidth]{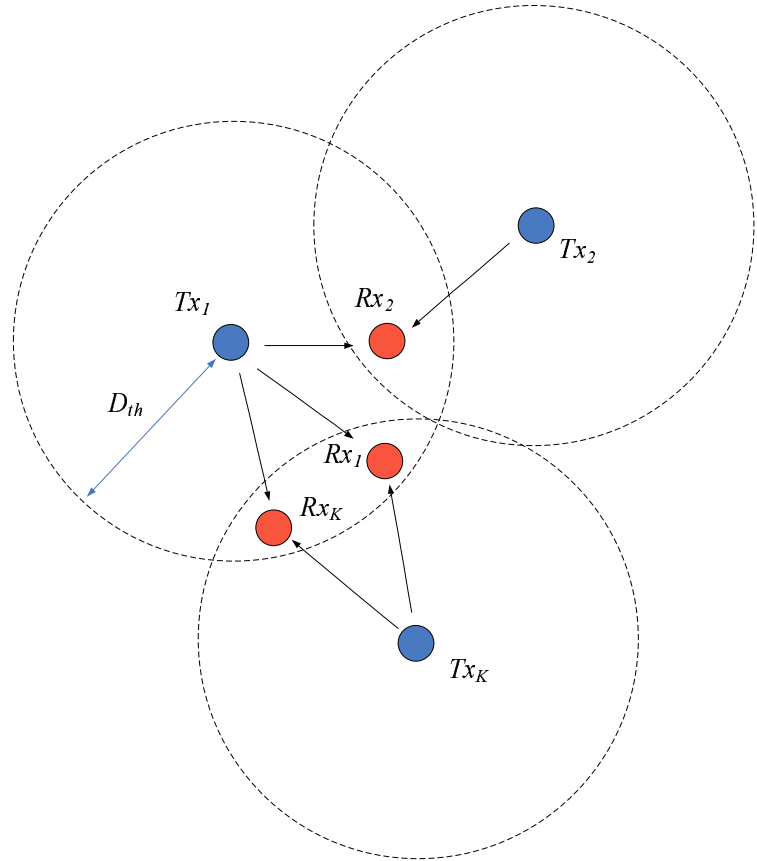}
\caption{System model of ad hoc networks.} \label{fg:adhoc}
\end{figure}

Consider a wireless ad-hoc network with a set
$\mathcal{K}=\{1,2,\ldots,K\}$ of distinct node pairs in Fig.
\ref{fg:adhoc}. Each link (node pair) consists of one dedicated
transmitter and one dedicated receiver. We assume that the
transmission of a link is interfered from the transmission of
another link, if the distance between the receiver node of the
former and the transmitter node of the latter is less than some
threshold $D_{th}$ \cite{Long}\cite{Mac_Num}. For any node $i$, we
define $I_i\subseteq\mathcal{K}$ as the set of nodes whose
transmitters cause interference to the receiver of node $i$ and
$O_i\subseteq\mathcal{K}$ as the set of nodes whose receivers get
interfered from the transmitter of node $i$. For example, in Fig.
\ref{fg:adhoc}, $I_1=\{K\} \ \textrm{and}\  O_1=\{2,K\}$. Then, the
throughput of node $i$ is
\begin{equation}
\label{eq:eqn42} u_k(\mathbf{p})=p_k\prod_{i \in I_k}(1-p_i).
\end{equation}
In this scenario, the state, namely, the contention measure signal,
can be redefined according to $s_k=\prod_{i \in I_k}(1-p_i)$.
Applying the conjecture-based approach, we have the following
conjectured utility function for node $k$:
\begin{equation}
\label{eq:eqn43} u_k^t(\tilde{s}_k^t(p_k),p_k)=p_k\Bigl[\prod_{i \in
I_k}(1-p_i^{t-1})-a_k(p_k-p_k^{t-1})\Bigr].
\end{equation}

Parallel to the theorems proven in Section III-B and C, we have the
following theorems on the stability and convergence of
conjecture-based bio-inspired learning algorithms in ad-hoc
networks. These theorems can be shown similarly as in Section III,
and hence, the proofs are omitted.

\subsubsection{Stability and Convergence}
\begin{theorem} \label{th:th10}
For any $\textbf{p}^*=(p_1^*,\ldots,p_K^*)\in P$, if
\begin{equation}
\sum_{i\in O_k\cup\{k\}}p_i^*<1, \ \forall k \in \mathcal{K}, \quad
\textrm{or} \quad \sum_{i \in I_k}\frac{p_k^*}{1-p_i^*}<1, \ \forall
k \in \mathcal{K},
\end{equation} $\textbf{p}^*$ is a stable CE for Algorithm 1 and Algorithm 2 with sufficiently small $\gamma$.
\end{theorem}
\begin{theorem} \label{th:th11}
Regardless of any initial value chosen for $\{p_k^0\}_{k=1}^{K}$, if
the parameters $\{a_k\}_{k=1}^{K}$ in the belief functions
$\{\tilde{s}_k\}_{k=1}^{K}$ satisfy
\begin{equation} \label{eq:eqn45}
a_k>|O_k|, \ \forall k \in \mathcal{K}, \quad \textrm{or} \quad
\sum_{i \in I_k}\frac{1}{a_i}<1, \forall k \in \mathcal{K},
\end{equation}
Algorithm 1 and Algorithm 2 with sufficiently small $\gamma$
converge to a unique CE.
\end{theorem}

\begin{remark} We observe that the sufficient conditions in Theorem
\ref{th:th10} and \ref{th:th11} are more relaxed compared with the
theorems in Section III. As opposed to the single-cell case, the
mutual interference is reduced in ad-hoc networks due to the large
scale geographical distance, therefore, these nodes can potentially
improve their throughput by increasing their transmission
probabilities while still maintaining the local stability as well as
global convergence.
\end{remark}

\begin{remark} In ad-hoc networks, the parameters $\{a_k\}_{k=1}^{K}$
can be determined in a distributed fashion such that the sufficient
conditions in Theorem \ref{th:th11} are satisfied. For example,
consider the symmetric case where transmitter $i$ interferes with
receiver $j$ if and only if transmitter $i$ can receive signals from
receiver $j$. Each transmitter can listen to the channel and
estimate $|O_k|$ by intercepting the ACK packets sent by the
receivers of the nodes in set $O_k$. An alternative distributed
solution is that each transmitter broadcasts its parameter $a_k$,
and receiver $k$ calculates $\sum_{i \in I_k}\frac{1}{a_i}$ and
notifies the nodes in set $I_k$ to adjust their parameters
accordingly.
\end{remark}

\subsubsection{Stability of the Throughput Region}

We also extend the stability analysis of the throughput region from
the single-cell scenario to the ad-hoc networks. The following lemma
explicitly describes the Pareto frontier of the throughput region.
\begin{lemma} \label{lm:lm12}
The Pareto boundary of the throughput region $\mathscr{T}$ can be
characterized as the set of points
$\pmb{\tau}=(\tau_1,\ldots,\tau_K)$ optimizing the weighted
proportional fairness objective \cite{Bell_lab}:
\begin{equation}\label{eq:eqn46}
\max_{\textbf{p} \in P} \sum_{k\in \mathcal{K}} \omega_k
\textrm{log} \tau_k,
\end{equation}
in which $\tau_k=p_k \prod_{i\in I_k}(1-p_i)$ for all possible sets
of positive link ``weights" $\{\omega_k\}_{k=1}^K$. Specifically,
for a particular weight combination $\{\omega_k\}_{k=1}^K$, the
optimal $\textbf{p}'$ is given by
\begin{equation}\label{eq:eqn47}
p_k'=\frac {\omega_k}{\omega_k+\sum_{i\in O_k} \omega_i}.
\end{equation}
\end{lemma}

\emph{Proof}: See \cite{Bell_lab} for details.

Based on Lemma \ref{lm:lm12}, we derive in the following theorem the
necessary and sufficient condition under which a particular
Pareto-efficient operating point is a stable CE for Algorithm 1.
Similar results can be derived for Algorithm 2 with sufficiently
small $\gamma$.

\begin{theorem} \label{th:th13}
Suppose $\textbf{p}^*=(p_1^*,\ldots,p_K^*)\in P$ satisfies
(\ref{eq:eqn47}) and maximizes the problem in (\ref{eq:eqn46}). The
elements of the Jacobi matrix $\textbf{J}$ at $\textbf{p}^*$ satisfy
\begin{equation} \label{eq:eqn48}
J_{ik}=\left\{ \begin{array}{cl}
\frac{1}{2},& \text{if $i=k$},\\
-\frac{p_i^*}{2(1-p_k^*)},& \text{if $k\in I_i$}, \\ 0,&
\text{otherwise}.
\end{array} \right.
\end{equation}
If $\rho(\textbf{J})<1$, $\textbf{p}$ is a stable CE for Algorithm
1.
\end{theorem}

\begin{remark} Theorem \ref{th:th13} generalizes the result in
Theorem \ref{th:th3} from the single-cell scenario to the ad-hoc
networks. Consider the $l_1$ norm for $\textbf{J}$ at $\textbf{p}$.
We have
\begin{equation} \label{eq:eqn49}
\lVert\textbf{J}\rVert_1=\max_{k \in \mathcal{K}} \frac
{\omega_k}{\omega_k+\sum_{i\in O_k}\omega_i}+ \sum_{i\in O_k}\frac
{\omega_i}{\omega_i+\sum_{j\in O_i}\omega_j}.
\end{equation}
In the single-cell case, $O_k=\mathcal{K}\backslash\{k\},\ \forall
k\in \mathcal{K}$, and $\lVert\textbf{J}\rVert_1$ equals to 1 for
any Pareto-optimal operating point. Therefore, any Pareto
inefficient operating point can be achieved with stability due to
$\rho(\textbf{J})\leq\lVert\textbf{J}\rVert_1<1$. However, in ad-hoc
networks, the form of the Jacobi matrix $\textbf{J}$ depends on the
actual network topology and it is difficult to bound the spectral
radius for a generic setting using certain matrix forms, such as
$l_1$ norm or $l_\infty$ norm. Alternatively, according to Theorem
\ref{th:th13}, we will numerically test the stability of the
Pareto-optimal operating points in the simulation section.
\end{remark}

\section{Numerical Simulations}
In this section, we numerically compare the performance of the
existing 802.11 DCF protocol, the P-MAC protocol \cite{fairness} and
the proposed algorithms in this paper.

\begin{figure}
\centering
\includegraphics[width=0.6\textwidth]{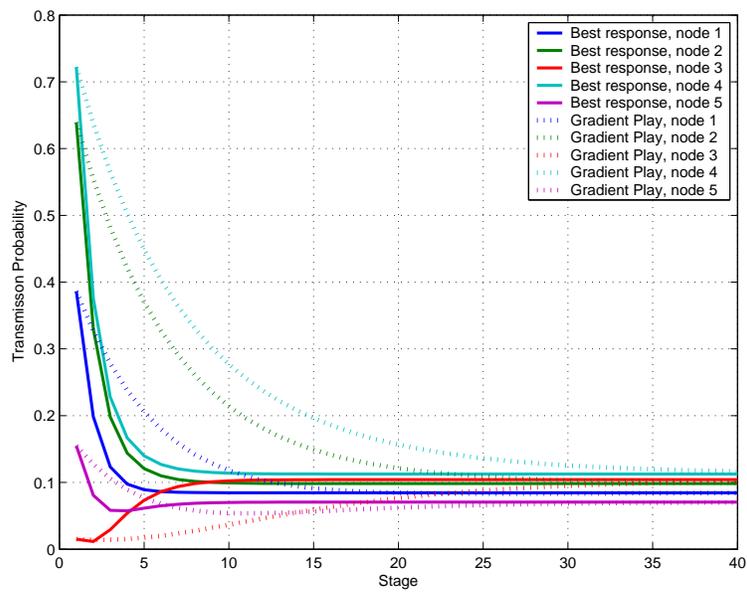}
\caption{Dynamics of Algorithms 1 and 2.} \label{fg:comparison}
\end{figure}

\begin{figure}
\centering
\includegraphics[width=0.6\textwidth]{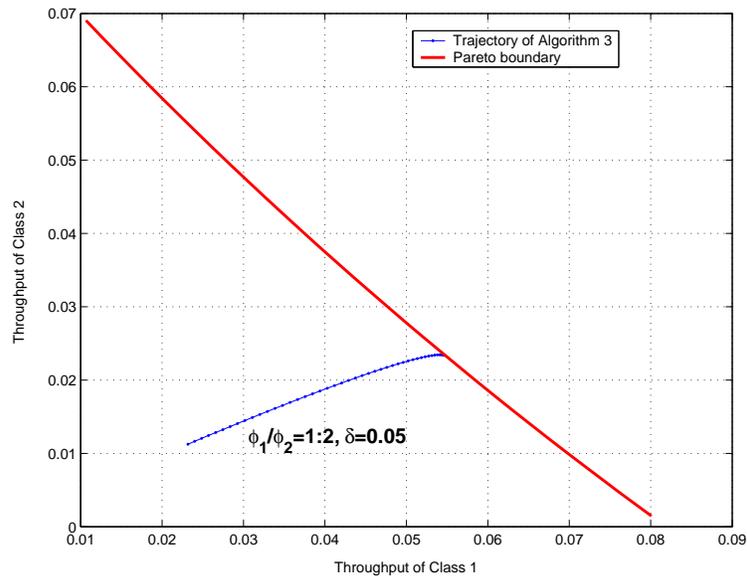}
\caption{The trajectory of Algorithm 3.} \label{fg:evolution}
\end{figure}

We first illustrate the evolution of transmission probabilities of
Algorithm 1 and 2. We simulate a single-cell network of 5 nodes. For
each node, the initial transmission probability $p_k^0$ is uniformly
distributed in $[0,1]$ and $a_k$ is uniformly distributed between 5
and 10. The stepsize in the gradient play is $\gamma=0.02$. Fig.
\ref{fg:comparison} compares the trajectory of the transmission
probability updates in both Algorithm 1 and 2 in a single
realization, under the assumption that node $k$ can perfectly
estimate the probability $\prod_{j \in \mathcal
{K}\backslash\{k\}}(1-p_j)$, $\forall k \in \mathcal {K}$. The best
response update converges in around 8 iterations and the gradient
play experiences a more smooth trajectory and the same equilibrium
is attained after 35 iterations. In addition, to illustrate how
individual nodes can adaptively adjust their algorithm parameters
and improve their throughput, we simulate a scenario with two
traffic classes. Each traffic class consists of 5 nodes and the
initial algorithm parameters of class 1 and 2 are $\phi_1=30$ and
$\phi_2=60$, respectively. The discount factor in Algorithm 3 is
$\delta=0.05$. The blue dotted curve in Fig. \ref{fg:evolution}
indicates that the operating point moves towards the red Pareto
boundary until the outer loop detects that the desired efficiency is
reached.


In practice, packet transmission over wireless links, e.g. IEEE
802.11 WLANs, involves extra protocol overheads, such as inter-frame
space and packet header. Assuming these realistic communication
scenarios, we compare various performance metrics, including
throughput, fairness, convergence, and stability, between our
proposed conjecture-based algorithms, the P-MAC protocol in
\cite{fairness}, and the IEEE 802.11 DCF. To evaluate these metrics,
the physical layer parameters need to be specified. In the
simulation, we assume that each wireless device operates at the IEEE
802.11a PHY mode-8, and the key parameters are summarized in Table
\ref{tb:table1}. We assume no transmission errors and the RTS/CTS
mechanism is disabled. The aggregate network throughput can be
calculated using Bianchi's model \cite{Bianchi}
\begin{equation} \label{eq:eqn50}
\mathcal{T}=\frac{P_s L_d}{(1-P_{tr})T_{slot}+P_s T_s+P_{tr}T_c-P_s
T_c},
\end{equation}
where $P_s=\sum_{n=1}^N|\mathcal{F}_n|\cdot p_n \cdot
(1-p_n)^{|\mathcal{F}_n|-1}\cdot \prod_{m\neq
n}(1-p_m)^{|\mathcal{F}_n|}$ is the probability that a transmission
occurring on the channel is successful,
$P_{tr}=1-\prod_{n=1}^N(1-p_n)^{|\mathcal{F}_n|}$ is the probability
that at least one transmission attempt happens, $T_s$ is the average
time of a successful transmission, and $T_c$ is the average duration
of a collision. The detailed derivation of $T_s$ and $T_c$ using the
given network parameters in Table \ref{tb:table1} can be found in
\cite{fairness}\cite{Bianchi}. The parameters in P-MAC are set
according to \cite{fairness}. The contention window sizes in the
IEEE 802.11 DCF are $CW_{min}=16$ and $CW_{max}=1024$. In Algorithm
3, individual nodes monitor the aggregate throughput to determine
whether to adjust the parameter $a_k$. The numerical results are
obtained using a MAC simulation program in \cite{Bianchi}. Our
comparison results are summarized as follows.

\begin{figure}
\centering
\includegraphics[width=0.6\textwidth]{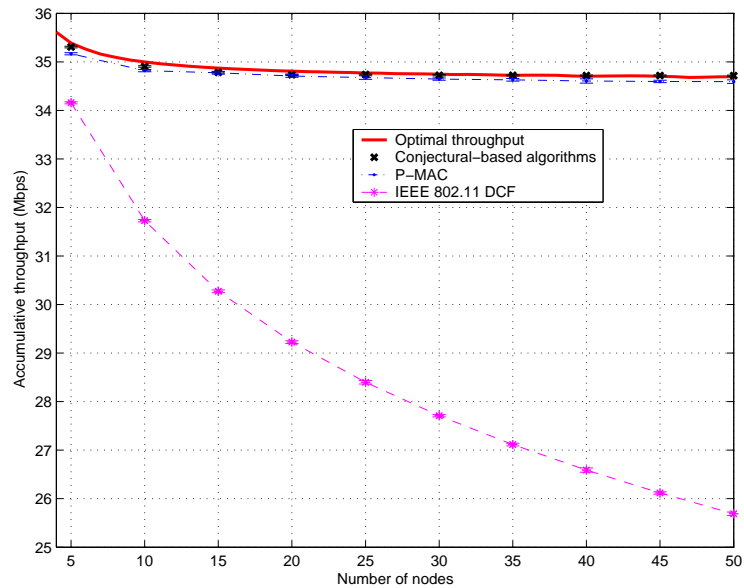}
\caption{Comparison of the accumulative throughput in the IEEE
802.11 DCF, P-MAC, and conjecture-based algorithms. Error bars
correspond to the standard deviation of the mean of the 100
measurements sampled at each point. The error bars in the remaining
figures are as in this figure.} \label{fg:throughput}
\end{figure}

First, the throughput of the three algorithms is compared. We vary
the total number of nodes $K$ from 4 to 50, in which $\lceil
K/2\rceil$ nodes carry class-1 traffic and the remaining nodes carry
class-2 traffic. The positive weights of class-1 and class-2 are
$\chi_1=1$ and $\chi_2=0.5$. The initial parameters in Algorithm 3
are chosen to be $\phi_1=3K/\chi_1$ and $\phi_2=3K/\chi_2$. As shown
in Fig. \ref{fg:throughput}, both the conjecture-based algorithm and
P-MAC significantly outperform the IEEE 802.11 DCF. The IEEE 802.11
DCF achieves the lowest throughput, because the lack of adaptation
mechanism of the contention window size causes more frequent packet
collisions as the number of nodes increases. Surprisingly, the
performance of the conjectural equilibrium attained by Algorithm 3
achieves the maximum achievable throughput. It also outperforms
P-MAC, because P-MAC uses approximation to derive closed-form
expressions for the transmission probabilities of different traffic
class.

\begin{figure}
\centering
\includegraphics[width=0.6\textwidth]{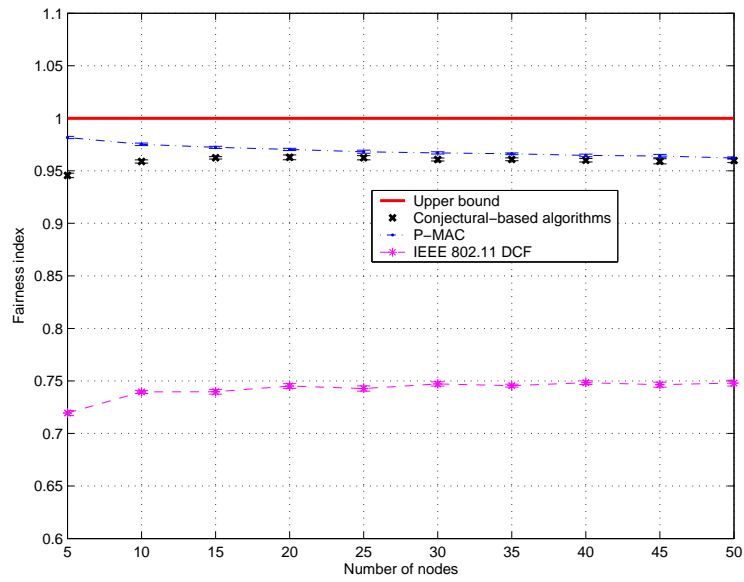}
\caption{Comparison of the achieved fairness of the IEEE 802.11 DCF,
P-MAC, and Algorithm 3.} \label{fg:fairness}
\end{figure}

Next, we evaluate the short-term fairness of different protocols
using the quantitative fairness index introduced in \cite{fairness}
\begin{equation} \label{eq:eqn51}
\textbf{F}=\frac{\mu(\mathcal{T}_k/\chi_n)}{\mu(\mathcal{T}_k/\chi_n)+\sigma(\mathcal{T}_k/\chi_n)},
k \in \mathcal{F}_n
\end{equation}
in which $\mathcal{T}_k$ denote the throughput of node $k$ that
belongs to traffic class $n$, and $\mu$ and $\sigma$ are,
respectively, the mean and the standard deviation of
$\mathcal{T}_n/\chi_n$ over all the active data traffic flows. We
simulate a transmission duration of 3 minutes. The stage duration in
Algorithm 3 is set as 50 successful transmissions. As shown in Fig.
\ref{fg:fairness}, we can see that Algorithm 3 and P-MAC are
comparable in their fairness performance and the achieved fairness
index is always above 0.95 regardless of the network configuration.
On the other hand, the fairness performance of 802.11 DCF is much
poorer than the previous two algorithms because the DCF protocol
provides no fairness guarantee.

\begin{figure}
\centering
\includegraphics[width=0.6\textwidth]{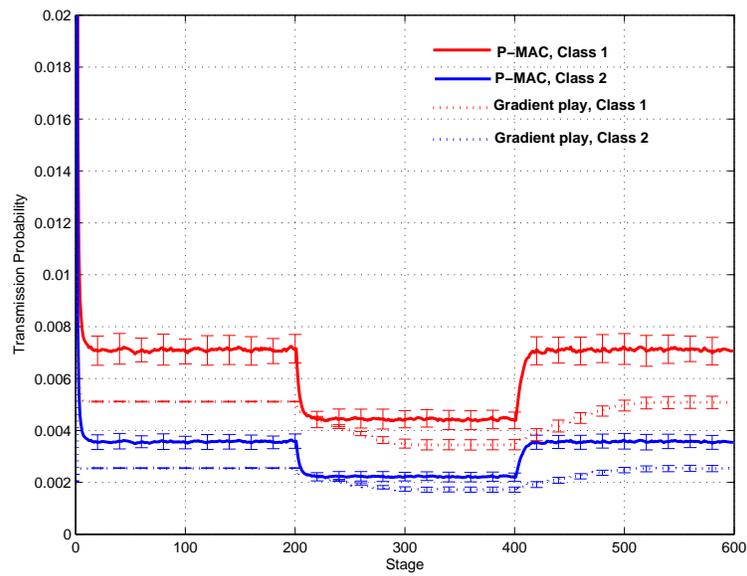}
\caption{The dynamics of the transmission probabilities in P-MAC and
Algorithm 3.} \label{fg:dynamics}
\end{figure}

\begin{figure}
\centering
\includegraphics[width=0.6\textwidth]{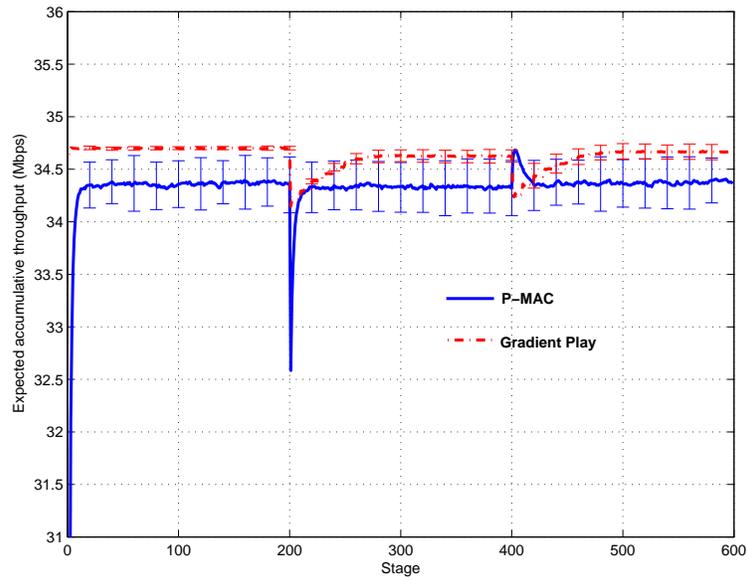}
\caption{The dynamics of the accumulative throughput in P-MAC and
Algorithm 3.} \label{fg:dynamics_throughput}
\end{figure}

Last, in order to compare the convergence and the stability of
different protocols for time-varying traffic, we simulate a network
in which the number of active nodes fluctuates over time. In order
to cope with traffic fluctuation, we slightly modify the outer loop
in Algorithm 3. Once some nodes join or leave the network (this can
be detected either by tracking the contention signal $\prod_{k \in
\mathcal{K}}(1-p_k)$ or estimating the total number of nodes in the
network \cite{Bianchi_est}), the adaptation of $a_k$ is activated.
Specifically, if more nodes join the network, $a_k\leftarrow
a_k(1+\delta)$, otherwise, $a_k\leftarrow a_k(1-\delta)$. At the
beginning, $|\mathcal{F}_1|=|\mathcal{F}_2|=25$. At stage 200, 15
class-1 and 15 class-2 nodes join the network. These nodes leave the
network at the 400th stage. The algorithm parameter $a_k$ is updated
every 5 stages and the stepsize in the gradient play is
$\gamma=0.003$. Fig. \ref{fg:dynamics} and Fig.
\ref{fg:dynamics_throughput} show the variation of the transmission
probabilities for both traffic classes and the expected accumulative
throughput over time. P-MAC does not converge due to the lack of
feedback control, which agrees with the observation about the
instability of P-MAC reported in \cite{Low_JSAC}. In addition, the
optimal transmission probabilities computed by P-MAC and the
conjecture-based algorithms are different under the same network
parameters because of the approximation used in P-MAC. As shown in
Fig. \ref{fg:dynamics}, nodes deploying P-MAC transmit with a higher
probabilities than the conjecture-based algorithms, which creates a
more congested environment. As a result, the accumulative throughput
achieved by P-MAC is slightly lower than the optimal throughput. In
contrast, the conjecture-based algorithms enable the nodes
adaptively tune their parameters $a_k$ to maximize the network
throughput while maintaining the weighted fairness as well as the
system stability. As shown in Fig. \ref{fg:dynamics} and Fig.
\ref{fg:dynamics_throughput}, during stage [200,300] and [400,470],
both the best response and the gradient play autonomously adapt
their parameter $a_k$ until it converges to the optimal operating
point. As discussed before, the best response learning converges
faster than the gradient play learning. To give a quantitative
measure of the stability, the standard deviations of the expected
accumulative throughput in Fig. \ref{fg:dynamics_throughput} for
different algorithms satisfy
$\sigma(\mathcal{T}_{P-MAC}^{Expected})/\sigma(\mathcal{T}_{BIO}^{Expected})\approx7$
and the actual achieved accumulative throughput satisfy
$\sigma(\mathcal{T}_{P-MAC}^{Actual})/\sigma(\mathcal{T}_{BIO}^{Actual})\approx2$.
We can see that, thanks to the inherent feedback control mechanism,
both bio-inspired learning algorithm exhibit superior stability
performance than P-MAC.

\begin{figure}
\centering
\includegraphics[width=0.6\textwidth]{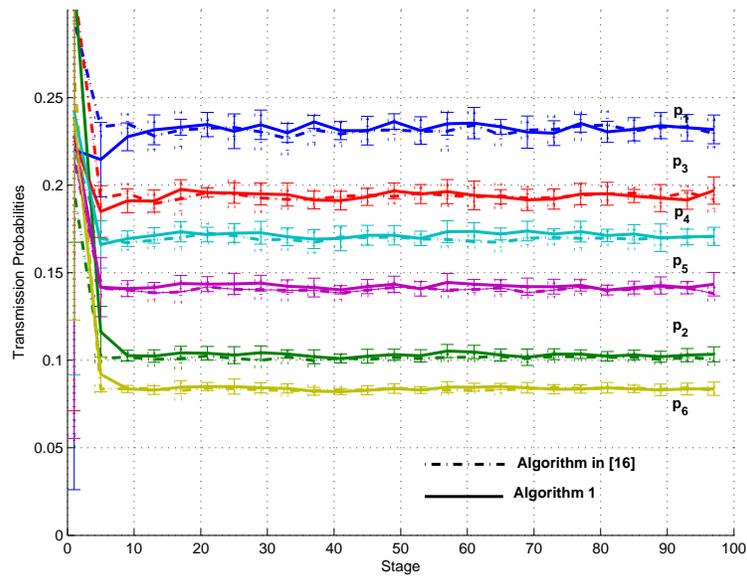}
\caption{Comparison between Algorithm 1 and the algorithm in
\cite{Num_nomsg}.} \label{fg:NumvsBio}
\end{figure}

We also simulate the evolution trajectory of the transmission
probabilities of the proposed Algorithm 1 and the algorithm in
\cite{Num_nomsg}. Both algorithms are essentially the best-response
based algorithms. Specifically, we consider a network with $K=6$.
The peak data rates for different nodes are $r_1=6$, $r_2=36$,
$r_3=9$, $r_4=12$, $r_5=18$, and $r_6=54$, all in Mbps. We apply the
algorithm in \cite{Num_nomsg} to solve the following network utility
maximization problem:
\begin{equation} \label{eq:eqn52}
\max_{\textbf{p}\in P}\sum_{k \in \mathcal {K}}
\frac{1}{1-\alpha}\bigl[ r_k p_k \prod_{j \in \mathcal
{K}\backslash\{k\}}(1-p_j)\bigr]^{1-\alpha},
\end{equation}
in which $\alpha=2$. The optimal solution corresponds to the belief
configuration $a_1=2.03, a_2=3.93, a_3=2.32, a_4=2.55, a_5=2.97$,
and $a_6=4.74$. The trajectory of both algorithms are shown in Fig.
\ref{fg:NumvsBio}. We can see that, both algorithms converge very
fast and oscillate around the neighborhood to the optimal solution
after several iterations. However, as we discussed before, the
algorithm in \cite{Num_nomsg} requires individual nodes to decode
all the received packet headers and estimate the transmission
probabilities of the other nodes individually, which introduces a
great internal computational overhead when the network size grows
large. In contrast, nodes deploying Algorithm 1 only have to
estimate the probability of having a free channel without the need
of decoding all the packets, which substantially reduces their
computational efforts.

\begin{figure}
\centering
\includegraphics[width=0.6\textwidth]{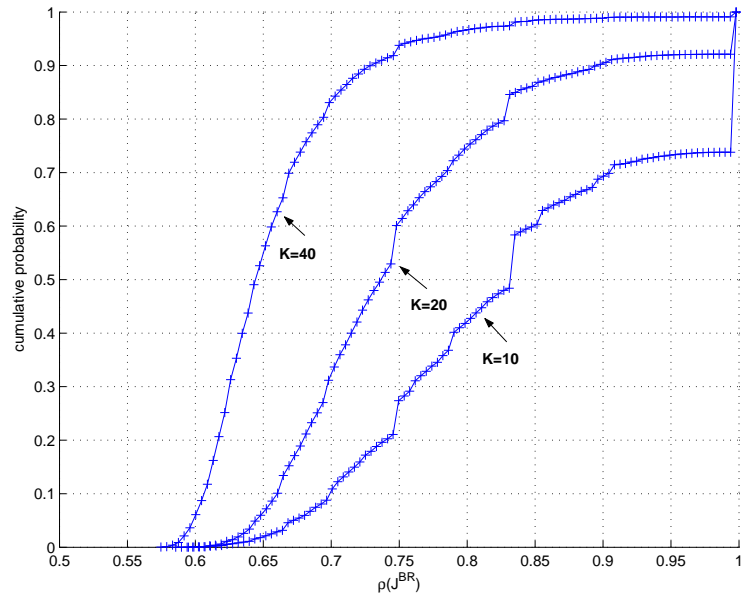}
\caption{Cumulative distribution function of $\rho(\textbf{J}^{BR})$
in ad-hoc networks.} \label{fg:ad_hoc}
\end{figure}

\begin{figure}
\centering
\includegraphics[width=0.6\textwidth]{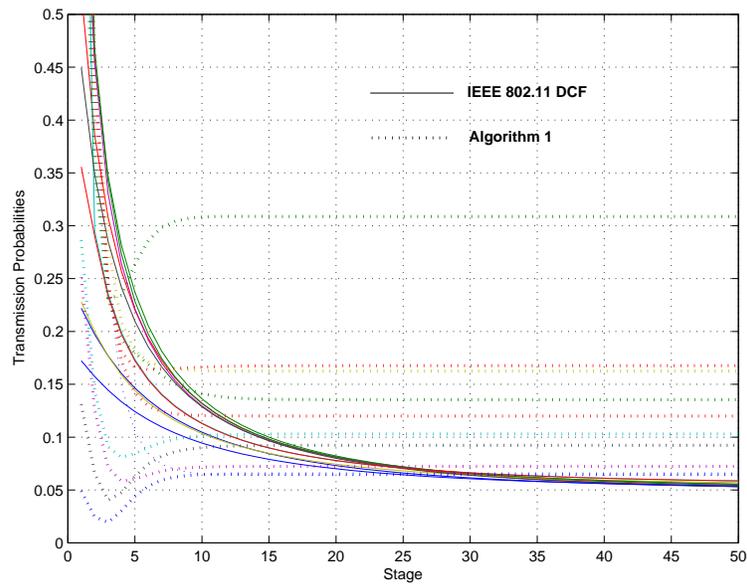}
\caption{Transmission probabilities of Algorithm 1 and the IEEE
802.11 DCF in ad-hoc networks.} \label{fg:adhoc_p}
\end{figure}

\begin{figure}
\centering
\includegraphics[width=0.6\textwidth]{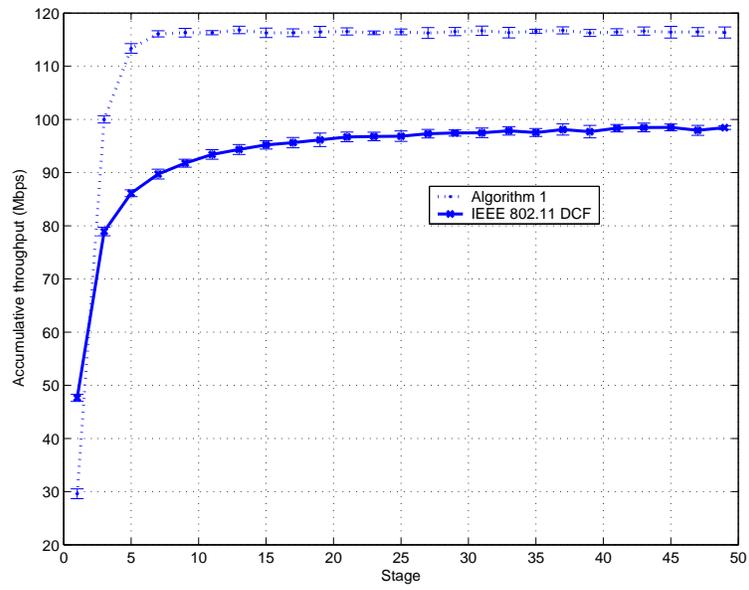}
\caption{Accumulative throughput of Algorithm 1 and the IEEE 802.11
DCF in ad-hoc networks.} \label{fg:adhoc_sumt}
\end{figure}

We simulate the performance of the proposed algorithms in an ad-hoc
network contained in a $100m\times100m$ square area. Nodes in the
square area are placed in the random manner. Two nodes can interfere
with each other if their distance is no more than $40m$, i.e.
$D_{th}=40m$. We simulate three scenarios with the node numbers
$K=\{10, 20, 40\}$. The Pareto-efficient point that we select is the
associated operating point with the link weighted vector
$\omega_k=1, 1\leq k \leq K/2, \textrm{and} \ 0.5, K/2 <k\leq K $ in
(\ref{eq:eqn47}). We can see from Fig. \ref{fg:ad_hoc} that,
$\rho(\textbf{J}^{BR})\leq1$ holds for all the simulated topologies.
As shown in Fig. \ref{fg:ad_hoc}, in some realizations,
$\rho(\textbf{J}^{BR})=1$, and hence, the associate operating points
are not asymptotically stable. This will occur when two nodes
interfere with each other and they do not interfere and are not
interfered by the remaining nodes in the entire ad-hoc network. On
the other hand, the stability improves as the number of nodes
increases. As long as the density of nodes is sufficiently large,
the stability of the conjecture-based algorithm on the
Pareto-efficient operating point can be achieved. Fig.
\ref{fg:adhoc_p} and Fig. \ref{fg:adhoc_sumt} show the evolution of
transmission probabilities and accumulative throughput for the IEEE
802.11 DCF and Algorithm 1 in a 10-node ad-hoc network with a
randomly generated topology. The trajectory of the IEEE 802.11 DCF
is obtained using the model in \cite{Rvs_Num}. The parameter $a_k$
in Algorithm 1 is chosen to be $|O_k|$. The intuition behind is
that, if $|O_k|=0$, node $k$ can transmit at the maximal probability
without interfering with any node. On the other hand, if $|O_k|$ is
large, node $k$ should backoff adequately such that the reciprocity
can be established. As shown in the figures, Algorithm 1 converges
faster and achieves higher throughput than DCF. Similar results have
been observed in the other simulated topologies.

\section{Conclusion}
In this paper, we propose distributed learning solutions that enable
autonomous nodes to improve their throughput performance in random
access networks. It is well-known that whenever biological entities
behave selfishly and myopically, a tragedy of commons might take
place, which has also been observed in the context of random access
control. Hence, we investigate whether forming internal belief
functions and learning the impact of various actions can alter the
interaction outcome among these intelligent nodes. Specifically, two
bio-inspired learning mechanisms are proposed to dynamically update
individual nodes' transmission probabilities. It is analytically
proven that the entire throughput region essentially consist of
stable conjectural equilibria. In addition, we prove that the
conjecture-based approach achieves the weighted fairness for
heterogeneous traffic classes and extend the distributed learning
solutions to ad-hoc networks. Simulation results have shown that the
proposed algorithms achieve significant performance improvement
against existing protocols, including the IEEE 802.11 DCF and the
P-MAC protocol, in terms of not only fairness and throughput but
also convergence and stability. A potential future direction is to
investigate how to detect and prevent misbehavior for these
bio-inspired solutions.


\begin{thebibliography}{200}


\bibitem{Haykin}
S. Haykin, ``Cognitive Radio: Brain-empowered wireless
communications", \emph{IEEE J. Selected Areas in Commu.}, vol. 23,
pp. 201-220, 2005.

\bibitem{Editor_1}
F. Dressler and I. Carreras, \emph{Advances in Biologically Inspired
Information Systems - Models, Methods, and Tools}, Studies in
Computational Intelligence (SCI), vol. 69, Berlin, Heidelberg, New
York, Springer, 2007.

\bibitem{Editor_2}
T. Suda, T. Itao, and M Matsuo, ``The Bio-Networking Architecture:
The Biologically Inspired Approach to the Design of Scalable,
Adaptive, and Survivable/Available Network Applications," in
\emph{The Internet as a Large-Scale Complex System}, the Santafe
Institute Book Series, Oxford University Press, 2005.

\bibitem{Gene}
R. Dawkins, \emph{The Selfish Gene}, 2nd ed., United Kingdom: Oxford
University Press, 1989.

\bibitem{Smith}
J. M. Smith, \emph{Evolution and the Theory of Games}, Cambridge:
Cambridge University Press, 1982.

\bibitem{learning_book}
D. Fudenberg and D. Levine, \emph{The Theory of Learning in Games},
Cambridge, MA: MIT Press, 1999.

\bibitem{Nowak}
M. Nowak, ``Five Rules for the Evolution of Cooperation,"
\emph{Science}, vol. 314, pp. 1560-1563, 2006.


\bibitem{Friedman}
E. Friedman and S. Shenker. ``Learning and Implementation on the
Internet", Manuscript. New Brunswick: Rutgers University, Department
of Economics, 1997. (available at
http://citeseer.ist.psu.edu/eric98learning.html)


\bibitem{Long}
C. Long, Q. Zhang, B. Li, H. Yang, and X. Guan, ``Non-Cooperative
Power Control for Wireless Ad Hoc Networks with Repeated Games",
\emph{IEEE J. Select. Areas Commun.}, vol. 25, pp. 1101-1112, Aug,
2007.

\bibitem{Fangwen}
F. Fu and M. van der Schaar,``Learning to Compete for Resources in
Wireless Stochastic Games", \emph{IEEE Trans. Veh. Tech.}, to
appear.





\bibitem{Aloha_eq}
Y. Jin and G. Kesidis, ``Equilibria of a noncooperative game for
heterogeneous users of an Aloha networks", \emph{IEEE Commun.
Lett.}, vol. 6, no. 7, pp. 282-284, July 2002.


\bibitem{Rvs_Num}
J. Lee, A. Tang, J. Huang, M. Chiang, and A. R. Calderbank,
``Reverse-engineering MAC: A non-cooperative game model", \emph{IEEE
J. Select. Areas Commun.}, vol. 25, no. 6, pp. 1135-1147, Aug. 2007.

\bibitem{Low_JSAC}
T. Cui, L. Chen, and S. H. Low, ``A Game-Theoretic Framework for
Medium Access Control", \emph{IEEE J. Select. Areas Commun.}, vol.
26, no. 7, pp. 1116-1127, Sep. 2008.

\bibitem{CSMACA_game}
\v{C}agalj, S. Ganeriwal, I. Aad, and J. P. Hubaux, ``On selfish
behavior in CSMA/CA networks", in \emph{Proc. IEEE Infocom}, pp.
2513-2524, Mar. 2005.

\bibitem{Mac_Num}
J. Lee, M. Chiang, and R. A. Calderbank, ``Utility-optimal
random-access control", \emph{IEEE Trans. Wireless Commun.}, vol. 6,
no. 7, pp. 2741-2751, July 2007.

\bibitem{Num_nomsg}
A. Mohsenian-Rad, J. Huang, M. Chiang, and V. Wong,
``Utility-Optimal Random Access: Optimal Performance Without
Frequent Explicit Message Passing", \emph{IEEE Trans. on Wireless
Commu.}, vol. 8, no. 2, 898-911, Feb. 2009.

\bibitem{G_Aloha}
R. T. B. Ma, V. Misra, and D. Rubenstein, ``An Analysis of
Generalized Slotted-Aloha Protocols", \emph{IEEE/ACM Trans.
Networking}, accepted for future publication.

\bibitem{Wellman}
M. P. Wellman and J. Hu, ``Conjectural equilibrium in multiagent
learning," \emph{Machine Learning}, vol. 33, pp. 179-200, 1998.

\bibitem{CE}
C. Figui\`{e}res, A. Jean-Marie, N. Qu\'{e}rou, and M. Tidball,
\emph{Theory of Conjectural Variations}, World Scientific
Publishing, 2004.

\bibitem{MIT_thesis}
S. G. Massaquoi, ``Modeling the function of the cerebellum in
scheduled linear servo control of simple horizontal planar arm
movements," Ph.D. dissertation, MIT, Cambridge, MA, 1999.

\bibitem{MIT_walking}
S. Jo, ``A neurobiological model of the recovery strategies from
perturbed walking," \emph{BioSystems}, vol. 90, no. 3, pp. 750-768,
2007.

\bibitem{gradient_1}
U. Dieckmann and R. Law, ``The dynamical theory of coevolution: a
derivation from stochastic ecological processes", \emph{Journal of
Mathematical Biology}, vol. 34, pp. 579-612, 1996.

\bibitem{gradient_2}
P. Taylor and T. Day, ``Evolutionary stability under the replicator
and the gradient dynamics", \emph{Evolutionary Ecology}, vol. 11,
pp. 579-590, 1997.

\bibitem{gradient_3}
L. Edelstein-Keshet, \emph{Mathematical Models in Biology}, Random
House, New York, 1988.

\bibitem{Learning_PB}
A. Jean-Marie and M. Tidball, ``Adapting behaviors through a
learning process", \emph{Journal of Economic Behavior and
Organization}, vol. 60, pp. 399-422, 2006.

\bibitem{Game_book}
R. Myerson, \emph{Game Theory}, Harvard University Press, 1991.


\bibitem{Matrix_book}
R. A. Horn and C. R. Johnson, \emph{Matrix Analysis}, Cambridge,
U.K. : Cambridge Univ. Press, 1991.

\bibitem{Contraction_mapping}
A. Granas and J. Dugundji, \emph{Fixed Point Theory}, New York:
Springer-Verlag, 2003.

\bibitem{AIMD}
D. Chiu and R. Jain, ``Analysis of the Increase/Decrease Algorithms
for Congestion Avoidance in Computer Networks", \emph{Journal of
Computer Networks and ISDN}, vol. 17, no. 1, pp. 1-14, June 1989.

\bibitem{IT_throughput}
J. Massey and P. Mathys, ``The collision channel without feedback",
\emph{IEEE Trans. Inform. Theory}, vol. 31, no. 2, pp. 192-204,
1985.

\bibitem{Bertsekas}
D. P. Bertsekas and J. N. Tsitsiklis, \emph{Parallel and Distributed
Computation}. Englewood Cliffs, New Jersey: Prentice Hall, 1997.


\bibitem{fairness}
D. Qiao and K. G. Shin, ``Achieving efficient channel utilization
and weighted fairness for data communications in IEEE 802.11 WLAN
under the DCF", \emph{Proc. IWQoS 2002}, pp. 227-236, May 2002.

\bibitem{Bell_lab}
P. Gupta and A. L. Stolyar, ``Optimal Throughput Allocation in
General Random-Access Networks", \emph{Proc. CISS 2006}, pp.
1254-1259, Mar. 2006.

\bibitem{802.11a}
IEEE 802.11a, \textit{Part 11: Wireless LAN Medium Access Control
(MAC) and Physical Layer (PHY) specifications: High-speed Physical
Layer in the 5 GHz Band}, Supplement to IEEE 802.11 Standard, Sep.
1999.

\bibitem{Bianchi}
G. Bianchi, ``Performance Analysis of the IEEE 802.11 Distributed
Coordination Function," \emph{IEEE J. Selected Areas in Commu.},
vol. 18, no. 3, pp. 535-547, Mar. 2000.

\bibitem{Bianchi_est}
G. Bianchi and I. Tinnirello, ``Kalman Filter Estimation of the
Number of Competing Terminals in an IEEE 802.11 Network,"
\emph{Proc. of IEEE Infocom}, 2003.

\bibitem{Mobicom_ref}
T. Nandagopal, T. E. Kim, X. Gao and V. Bharghhavan, ``Achieving MAC
layer fairness in wireless packet networks," \emph{Proc. ACM
Mobicom}, 2000.

\bibitem{PID}
K. J. Astrom and T. Hagglund, \emph{PID Controllers: Theory, design
and tuning}. Instrument society of America, 2nd edition, 1995.

\bibitem{SM}
J. S. Simonoff, \emph{Smoothing Methods in Statistics, Springer
Series in Statistics}. Springer-Verlag New York, 1996.

\end{thebibliography}

\newpage

\begin{table}[h]
\caption{IEEE 802.11a PHY mode-8 parameters}
\begin{center}
\begin{tabular}{|c|c|}
\hline
Parameters & Value \\
\hline \hline
Duration of an Idle Slot ($T_{slot}$) & 9 $\mu s$ \\
Duration of PHY Header ($T_{PHY}$) & 20 $\mu s$ \\
SIFS Time ($T_{SIFS}$) & 16 $\mu s$ \\
DIFS Time ($T_{DIFS}$) & 34 $\mu s$ \\
Propagation Delay ($T_{d}$) & 1 $\mu s$ \\
MAC Header ($L_{MAC}$) & 28 octets \\
Packet Payload Size ($L_{d}$) & 2304 octets \\
ACK Frame Size ($L_{ACK}$) & 14 octets \\
Data Rate ($R_{t}$) & 54 Mbps \\
\hline
\end{tabular}
\end{center}
\label{tb:table1}
\end{table}

\end{document}